\documentclass{aa}
\usepackage{here}

\usepackage{natbib}
\title{ Echoes of multiple outbursts of Sagittarius A$^{\star}$ \\ revealed by Chandra}
\titlerunning{Echoes of multiple outbursts of Sagittarius A$^{\star}$ revealed by Chandra}

\author{M. Clavel\inst{\ref{instAPC}\and\ref{instCEA}} \and R. Terrier\inst{\ref{instAPC}}\and A. Goldwurm\inst{\ref{instAPC}\and\ref{instCEA}} \and M. R. Morris\inst{\ref{UCLA}} \and G. Ponti\inst{\ref{MPE}} \and S. Soldi\inst{\ref{instAPC}} \and G. Trap\inst{\ref{palais}\and\ref{instAPC}\and\ref{instCEA}} }
\institute{AstroParticule et Cosmologie, Universit\'e Paris Diderot, CNRS/IN2P3, CEA/DSM, Observatoire de Paris, Sorbonne Paris Cit\'e ; 10, rue Alice Domon et L\'eonie Duquet, 75205 Paris Cedex 13, France\label{instAPC}
\and Service d'Astrophysique/IRFU/DSM, CEA Saclay, Bat. 709, 91191 Gif-sur-Yvette Cedex, France\label{instCEA}
\and Department of Physics \& Astronomy, University of California, Los Angeles, CA 90095-1547, USA\label{UCLA}
\and Max-Planck-Institute for Extraterrestrial Physics, Garching, PSF 1312, D-85741 Garching, Germany\label{MPE}
\and Unit\'e de physique, Palais de la d\'ecouverte - Universcience, 75008 Paris, France  \label{palais}
}
\authorrunning{Clavel et al.}
\date{Received 8 April 2013 / Accepted 12 July 2013} % / Accepted ...

\def\sgr{$\rm Sgr \, A^{\star}$}
\def\fe{Fe~K$\alpha$}
\def\nh{N$_H$}

\begin{document}

%%%%%%%%%%%%%%%%%%%%%%%%%%%%%%%%%%%% ABSTRACT and KEYWORDS
\abstract
{
The relatively rapid spatial and temporal variability of the X-ray radiation from some molecular clouds near the Galactic center shows that this emission component is due to the reflection of X-rays generated by a source that was luminous in the past, most likely the central supermassive black hole, Sagittarius~A$^{\star}$.
}
{
Studying the evolution of the molecular cloud reflection features is therefore a key element to reconstruct \sgr's past activity. The aim of the present work is to study this emission on small angular scales in order to characterize the source outburst on short time scales.
}
{
We use Chandra high-resolution data collected from 1999 to 2011 to study the most rapid variations detected so far, those of clouds between 5$'$ and 20$'$ from \sgr\ towards positive longitudes.
Our systematic spectral-imaging analysis of the reflection emission, notably of the \fe\ line at 6.4~keV and its associated 4--8~keV continuum, allows us to characterize the variations down to 15$''$ angular scale and 1-year time scale.
}
{
We reveal for the first time abrupt variations of few years only and in particular a short peaked emission, with a factor of 10 increase followed by a comparable decrease, that propagates along the dense filaments of the ``Bridge'' cloud. 
This 2-year peaked feature contrasts with the slower 10-year linear variations we reveal in all the other molecular structures of the region. 
Based on column density constraints, we argue that these two different behaviors are unlikely to be due to the same illuminating event.
}
{
The variations are likely due to a highly variable active phase of \sgr\ sometime within the past few hundred years, characterized  by at least two luminous outbursts of a few-year time scale and during which the \sgr\ luminosity went up to at least $10^{39}$ erg s$^{-1}$.
}


\maketitle
%
%
%
%
%
%
%
%%%%%%%%%%%%%%%%%%%%%%%%%%%%%%%%%%%% INTRO
\section{Introduction}
\sgr\ is the supermassive black hole located in the center of the Galaxy, at a distance of about 8~kpc. Its estimated mass is $4\times10^6$~M$_\odot$ \citep{ghez2008,gillessen2009a}, and despite the presence of gaseous features and stellar winds in the near environment surrounding the black hole, its quiescent X-ray luminosity is only about $10^{33-34}$~erg~s$^{-1}$ \citep{baganoff2003}. This has motivated the development of several theoretical models based on radiatively inefficient accretion flows \citep{Melia2001}. 
The intensity of \sgr\ is also known to vary, showing rapid flares during which its luminosity has been observed to increase by a factor up to 160 \citep[][and references therein]{nowak2012}. Nevertheless, the current activity of \sgr\ remains at least eight orders of magnitude lower than its Eddington luminosity, making this specimen one of the least luminous known supermassive black holes. 
The recent detection of a dense gas cloud falling towards the accretion zone of \sgr\ \citep{gillessen2012, gillessen2013} provides evidence that the accretion rate onto \sgr\ may vary, and since Active Galactic Nuclei (AGN) have a short duty cycle \citep[$\sim10^{-2}$,][]{greene2007}, \sgr\ is not incompatible with being a low luminosity AGN in a temporary low state. \\\\
Furthermore, there are strong hints that \sgr\ has experienced a higher level of activity in the past \citep[see][for a review]{ponti2012}. The large Fermi Bubbles observed in the GeV energy range and extending 10 kpc above and below the Galactic center \citep{su2010} could be the vestiges of such past activity. Indeed, even if it is not the only possible explanation, a past accretion event onto \sgr\ could have injected enough energy into the Galactic center to create such structures in the last ten million years. On shorter time scales, the recent history of \sgr\ can be reconstructed from the non-thermal emission emanating from the molecular clouds at the Galactic center \citep{sunyaev1993, koyama1996, murakami2000, revnivtsev2004, muno2007, inui2009, ponti2010, ponti2012, terrier2010, capelli2011, capelli2012, nobukawa2011, ryu2012}. 
\\\\  
The Galactic Central Molecular Zone \citep[CMZ,][]{morris1996}, which is composed of massive molecular clouds in the inner core of the Galaxy, displays strong and diffuse X-ray emission. Between 2 and 10~keV, this complex emission is at least composed of a uniformly distributed soft component described as a low temperature plasma, a less uniform but centrally peaked 6.7~keV line modelled by a hot plasma and clumpy 6.4~keV iron line emission correlated with molecular structures \citep{park2004, nobukawa2010}.  The presence of the X-ray fluorescent line of neutral iron was predicted by \citet{sunyaev1993} and first detected by \citet{koyama1996}. 
The strong variability of the 6.4 keV emission, detected in both Sgr~B2 \citep{inui2009, terrier2010} and the Sgr~A region \citep{ponti2010, capelli2012}, proves that an important fraction of the diffuse non-thermal emission is due to reflection. This reflected emission is created by Compton scattering and K-shell photo-ionization of neutral iron atoms produced by an intense X-ray radiation such as could have  originated as a past large outburst of \sgr\ \citep{sunyaev1993, koyama1996, sunyaev1998}. Nevertheless, the attribution of a specific emission feature to this physical process is not trivial since it can also be produced by the interaction of energetic charged particles, either fast electrons \citep{yusef2002}, subrelativistic protons \citep{dogiel2009}, or nuclei \citep{tatischeff2012}, with molecular clouds.
\\\\
Using XMM-Newton data from 2000 to 2009, \citet{ponti2010} presented a characterization of Sgr~A illumination variations with the first detection of superluminal apparent propagation in the collection of molecular clouds located between \sgr\ and the Radio Arc. Since molecular cloud properties and positions along the line of sight are poorly constrained in this region, conclusions about the past activity of \sgr\ are not straightforward. In particular, molecular clouds at different projected distances do not necessarily imply different flares from \sgr.
\citet{ponti2010} showed that the data available up to 2009 was still consistent with just one single long illuminating event fading about 100 years ago. According to this scenario, the luminosity of \sgr\ might have been around $10^{39}$~erg~s$^{-1}$ for a few centuries. 
The lightcurve of \sgr's past luminosity could also be more complex, as suggested by an alternative geometry derived by \citet{capelli2012} for the Sgr~A region. Using Suzaku data on Sgr~B and Sgr~C, \citet{ryu2012} also reported a flux variability of \sgr\ during the past few centuries and interpreted it as multiple flares superposed upon a long-term high-state activity of \sgr.\\\\
\sgr's past higher level of activity seems manifest but the precise structure of this activity is less clear. The main focus of this work is to study both the variable emission in the \fe\ and in the continuum (4--8~keV) in the key region located between \sgr\ and the Radio Arc (hereafter the Sgr~A complex), where the strongest variations have been detected. In particular, we take advantage of Chandra high spacial resolution in order to highlight the fine structure in the variable illumination.\\\\
Section 2 presents the observations and the data reduction. The analyses of the variations are then divided into three sections according to the scale and the energy range used. Section~3 describes a 6.4~keV map of the Sgr~A complex and a three-period RGB map in order to localize the non-thermal emission and to provide the overall picture of its variations. Section~4 presents the \fe\ emission variations of two clouds having very different time behaviors. Section~5 presents a systematic 4--8~keV  analysis of the small-scale variations over a large part of the Sgr A complex. It gives the spatial distribution of the two different time behaviors represented in the complex. In section~6, the possible origins of the different observed time behaviors are discussed and interpreted in terms of two past outbursts of \sgr. The conclusions are summarized in section~7.
%
%
%
%
%
%%%%%%%%%%%%%%%%%%%%%%%%%%%%%%%%%%%% DATA
\section{Observations and Data Reduction}
In order to follow the interstellar echo of the past activity of \sgr, we were granted a specific 160~ks observation run with Chandra in July 2011. Our analysis is focussed on the ten arcmin squared region centred on (l,b)$\,=\,$($0.06^\circ$,~$-0.10^\circ$). To understand the pattern of its variations we also use all available Chandra data of this region from 1999 to 2010, restricting our analysis to the ACIS-I data in order to avoid background and off centred point spread function issues.
All observations used in this work are detailed in Table \ref{tab:obsID}. Since a large exposure time is needed to perform a precise analysis, we group the different pointings according to the observation year.
\begin{table*}
	\centering
	\caption{Chronological observation number,  observation starting date, total observation time, cleaned exposure time, Chandra observation identification, satellite pointing in galactic coordinates (\sgr\ is at (l,b)$\,=\,$($359.944^\circ$,~$-0.046^\circ$)).}
	\label{tab:obsID}
	\begin{scriptsize}	
	\begin{tabular*}{0.9\textwidth}{@{\extracolsep{\fill}}r l r r r c}
	\hline \hline
	N & Date & Time & Exposure & Obs. ID & Pointing\\
	  &      & (ks) & (ks) &         & (l,b)\\
	\hline
	1 & 1999 Sept 21 & 46.5 & 45.0 & 242 & \sgr \\
	\hline
	2 & 2000 Jul 07 & 49.4 & 46.8 & 945 & (0.141, -0.097)\\
	3 & 2000 Oct 26 & 50.0 & 11.4 & 1561 & \sgr \\
	\hline
	4 & 2001 Jul 18 & 11.8 & 11.5 & 2273 & (0.195, -0.195)\\
	5 & 2001 Jul 18 & 11.8 & 11.3 & 2276 & (0.195, 0.000)\\
	6 & 2001 Jul 18 & 10.8 & 10.5 & 2282 & (0.000, -0.195)\\
	7 & 2001 Jul 18 & 10.8 & 10.5 & 2284 & (360.000, 0.000)\\
    \hline 
	12 & 2002 May 22 & 38.5 & 29.7 & 2943 & \sgr \\ 
	8  & 2002 Feb 19 & 12.5 & 12.4 & 2951 & \sgr \\ 
	9  & 2002 Mar 23 & 12.0 & 11.4 & 2952 & \sgr \\ 
	10 & 2002 Apr 19 & 11.9 & 10.2 & 2953 & \sgr \\ 
	11 & 2002 May 07 & 12.6 & 12.0 & 2954 & \sgr \\ 
	14 & 2002 May 25 & 168.9 & 160.7 & 3392 & \sgr \\
	15 & 2002 May 28 & 160.1 & 156.6 & 3393 & \sgr \\
	13 & 2002 May 24 & 38.5 & 33.9 & 3663 & \sgr \\
	16 & 2002 Jun 03 & 91.1 & 89.3 & 3665 & \sgr \\
	\hline
	17 & 2003 Jun 19 & 25.1 & 24.7 & 3549 & \sgr \\
	\hline
	18 & 2004 Jun 09 & 99.8 & 97.0 & 4500 & (0.122, 0.019) \\
	19 & 2004 Jul 05 & 50.1 & 49.3 & 4683 & \sgr \\
	20 & 2004 Jul 06 & 50.1 & 49.0 & 4684 & \sgr \\
	21 & 2004 Aug 28 & 5.2 & 4.9 & 5360 & \sgr \\
	\hline
	23 & 2005 Jul 24 & 49.4 & 48.3 & 5950 & \sgr \\
	24 & 2005 Jul 27 & 46.4 & 44.6 & 5951 & \sgr \\
	25 & 2005 Jul 29 & 46.3 & 44.3 & 5952 & \sgr \\
	26 & 2005 Jul 30 & 46.0 & 35.9 & 5953 & \sgr \\
	27 & 2005 Aug 01 & 18.3 & 16.7 & 5954 & \sgr \\
	22 & 2005 Feb 27 & 4.9 & 4.6 & 6113 & \sgr \\
	\hline
	32 & 2006 Jul 17 & 30.2 & 29.3 & 6363 & \sgr \\
	28 & 2006 Apr 11 & 4.6 & 2.5 & 6639 & \sgr \\
	29 & 2006 May 03 & 5.2 & 4.8 & 6640 & \sgr \\
	30 & 2006 Jun 01 & 5.1 & 4.9 & 6641 & \sgr \\
	31 & 2006 Jul 04 & 5.2 & 5.0 & 6642 & \sgr \\
	33 & 2006 Jul 30 & 5.1 & 4.9 & 6643 & \sgr \\
	34 & 2006 Aug 22 & 5.1 & 4.6 & 6644 & \sgr \\
	35 & 2006 Sep 25 & 5.2 & 4.2 & 6645 & \sgr \\
	36 & 2006 Oct 29 & 5.2 & 4.3 & 6646 & \sgr \\
	\hline
	37 & 2007 Feb 14 & 38.7 & 38.0 & 7048 & (0.184, -0.199)\\
	38 & 2007 Feb 11 & 5.2 & 4.8 & 7554 & \sgr \\
	39 & 2007 Mar 25 & 5.2 & 4.9 & 7555 & \sgr \\
	40 & 2007 May 17 & 5.0 & 4.8 & 7556 & \sgr \\
	41 & 2007 Jul 20 & 5.1 & 4.7 & 7557 & \sgr \\
	42 & 2007 Sep 02 & 5.0 & 4.7 & 7558 & \sgr\ \\
	43 & 2007 Oct 26 & 5.1 & 4.9 & 7559 & \sgr \\
	\hline
	44 & 2008 May 05 & 28.0 & 27.5 & 9169 & \sgr \\
	45 & 2008 May 06 & 27.2 & 26.4 & 9170 & \sgr \\
	46 & 2008 May 10 & 28.0 & 27.5 & 9171 & \sgr \\
	47 & 2008 May 11 & 27.8 & 27.3 & 9172 & \sgr \\
	49 & 2008 Jul 26 & 28.1 & 27.4 & 9173 & \sgr \\
	48 & 2008 Jul 25 & 29.2 & 28.0 & 9174 & \sgr \\
	\hline
	50 & 2009 May 18 & 114.0 & 110.6 & 10556 & \sgr \\
	\hline
	51 & 2010 May 13 & 80.0 & 78.5 & 11843 & \sgr \\
	\hline
	53 & 2011 Jul 21 & 59.3 & 56.1 & 12949 & (0.013, -0.088)\\
	54 & 2011 Jul 29 & 67.1 & 64.8 & 13438 & (0.013, -0.088)\\
	52 & 2011 Jul 19 & 31.9 & 30.6 & 13508 & (0.013, -0.088)\\
	\hline \hline
	\end{tabular*}
	\end{scriptsize}	
\end{table*}
Most of the selected observations are pointed towards \sgr\ and, due to Chandra field of view and to the different observation strategies, the eastern part of the region is not uniformly covered.
\\\\
We use the latest version of the data available in the Chandra archive as of July 1st 2012, which includes  up-to-date astrometry and energy calibration.
The data reduction is then performed using Chandra software CIAO version~4.4{\footnote{see http://cxc.harvard.edu/ciao/index.html for more details.}}.
\subsection{Flux Mosaics}
Bad pixels are removed using the CIAO \textit{ardlib} routine and \textit{bpix1.fits} file provided with the data. Using the \textit{celldetect} routine, we detect point sources and exclude
them to extract the observation lightcurve in order to detect particle background flares. We cut such time intervals by running twice the \textit{deflare} routine, which removes events with count rate higher than twice the total mean count rate.\\\\
To estimate the background contribution, Chandra analysis software provides blank sky event files. We reproject them in order to match the observation astrometry using the \textit{reproject\_{}events} routine and the \textit{asol1.fits} file provided with the data. In this work they are used as background event files.\label{sec:ch-bkg}
Instrumental lines and Galactic low-energy emission are not traced by this background estimator.
Therefore, following \citet{ponti2010}, we restrict our analysis to the 4--8~keV energy range in order to have a relevant estimate of the background. 
\\\\
In the 4--8~keV range, the strongest signature of the high-energy reflection is the \fe\ fluorescent line at 6.4~keV \citep{nandra1994}. To visualize its variation we first build continuum-subtracted \fe\ flux maps. For each event file, counts with energy between 6.32 and 6.48~keV are integrated over time and the corresponding exposure map is created using the \textit{merge\_{}all} routine. The same procedure is applied to counts with energy in the 4--6.32 and 6.48--8~keV bands. Then, assuming the continuum spectral shape is described by a power-law of photon index $\Gamma=2$, we rescale this continuum image (using a factor of 0.052) to obtain the contribution of the continuum emission underlying the \fe\ line.
If several observations are grouped, signal, continuum and exposure maps are merged into their respective mosaics using the \textit{reproject\_{}image\_{}grid} routine. Then, the final flux mosaic is obtained by dividing the signal mosaic by the corresponding exposure mosaic and subtracting the associated continuum flux.
\subsection{Spectral analysis} 
In order to better constrain the \fe\ line variability we also study the spectral shape of the emission by fitting it with a simple model. In order to obtain the required statistics, this analysis is restricted to the following nine years: 2000, 2002, 2004, 2005, 2006, 2008, 2009, 2010 and 2011.\\\\
For each observation, we use the \textit{specextract} routine to extract source and background spectra for extended regions and to build the associated weighted Ancillary Response File (ARF) and Redistribution Matrix File (RMF).
Because most observations do not have sufficient counts to conduct a proper analysis on the region we are interested in, we sum spectra per year and region using the \textit{combine\_{}spectra} routine. As the energy calibration depends on the position within a CCD, it varies from observation to observation, therefore the spectra summation slightly increases the \fe\ line width and prevents us from using this parameter in the analysis. 
Lastly, to improve the performance of the spectral fit, spectrum counts are grouped with \textit{group\_{}counts} to have at least forty counts per bin.\\\\ 
We choose to restrict the spectral energy range to \hbox{4--7.1~keV} because the reflection component contribution is larger in this range and the soft plasma (1 keV) contribution is reduced. The Chandra data statistics does not allow a full characterization of the spectra, but previous analysis of the region with both XMM-Newton \citep{ponti2010,capelli2012} and Suzaku \citep{nobukawa2010} demonstrated that they are compatible with the superposition of a reflection spectrum and a thermal component. 
The simplest model fitting the derived spectra is therefore a gaussian \fe\ emission line added to a power-law continuum and to a thermal emission characterized by a 6.5~keV temperature plasma \citep{koyama2007}. The poor statistics and the energy range are limits to properly determine both the photon index and the column density parameters of the absorbed power-law, so we decided to use a simple power-law to model the continuum emission even if its index cannot be considered at face value.
Therefore, we fit the data using the Xspec model: \textsc{gaussian~$+$~powerlaw~$+$~apec}. For each region, the hot-plasma component is expected to be constant over the years whereas the reflection of high-energy emission induces the \fe\ line and the continuum emissions to vary in a correlated way. In order to reduce the number of free parameters, the hot-plasma temperature is fixed to 6.5~keV \citep{koyama2007} and the same normalization is used across the years for a given region. %
\label{sec:DA-specfit}
Moreover, the iron line width and position are assumed to be constant across all studied regions for a given year. %
Hence, we first fit these two parameters for each year on a large region spectrum, these values being then fixed for all fits on individual regions.
Therefore, for each region, we fit simultaneously the $9\times3+1$ free parameters left on the nine year spectra coverage. The free parameters are the hot-plasma normalisation, the nine years iron line amplitudes and the nine years continuum indices and normalisations. The fit is performed using a chi-square statistic with the Gehrels variance function \citep{gehrels1986}.
\label{sec:DA-lc}
For regions large enough to conduct a proper spectral analysis, we build \fe\ flux lightcurves.
The errors on the flux are given by the confidence interval of the fit at~$1\,\sigma$.
%
%
%
%
%%%%%%%%%%%%%%%%%%%%%%%%%%%%%%%%%%%% GLOBAL
\section{\fe\ emission: global view of the central 30~pc}
The diffuse emission due to reflection of high-energy radiation ($>\,7.1$~keV) is mainly characterized by the neutral iron line emission at 6.4 keV. We identify the large areas emitting at this energy and characterize their overall \fe\ variations in this section.
\subsection{Correlation with molecular structures}
To visualize the distribution of neutral Fe K emission in the Sgr A complex we build a mosaic map based on available Chandra data from 1999 to 2011. As the large amount of 2002 exposure time tends to dominate the flux mosaic, we decided to use only observation 3392 for this year, in order to properly identify all the regions emitting within the period 1999--2011.
\begin{figure*}
	\centering
	\includegraphics[trim = 0mm 0mm 10mm 0mm,clip,width=1\textwidth]{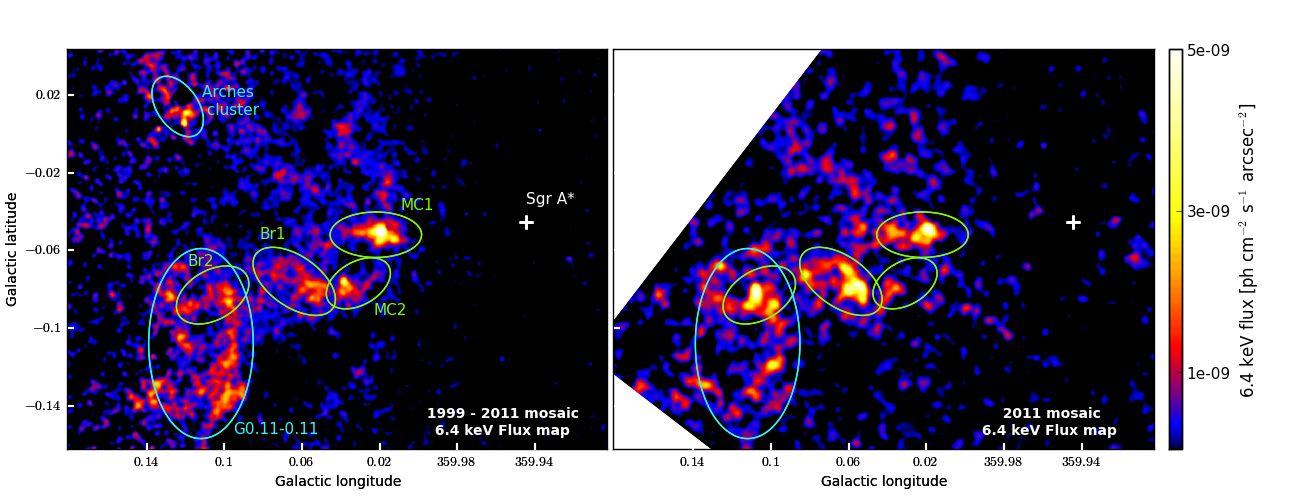}
	\caption{Chandra continuum subtracted flux maps of the neutral Fe~K emission line at 6.4~keV. \textit{(Left)} 1999 - 2011 mosaic smoothed using a 4~arcsec gaussian kernel (only observation 3392 has been kept among 2002 observations to keep the time coverage as uniform as possible). The emission of the region is mainly distributed in five distinct features named MC1, MC2, Br1, Br2 and G0.11-0.11 but fainter structures are also visible. \textit{(Right)} 2011 mosaic smoothed to 6~arcsec because of the lower statistics on this second map. MC1, Br1 and Br2 are very bright while the other two clouds are faint.}
	\label{fig:ChandraView}
\end{figure*}
\begin{figure*}
	\centering
	\includegraphics[trim = 0mm 2mm 0mm 25mm,clip,width=1\textwidth]{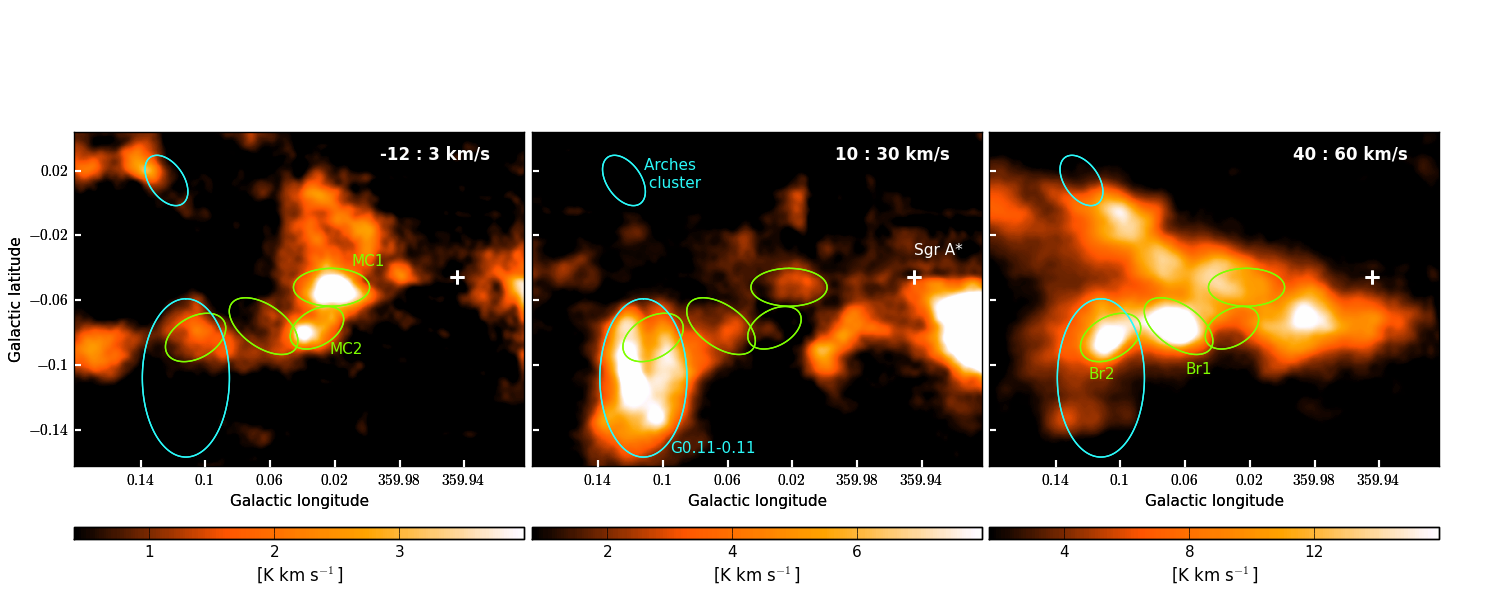}
	\caption{ Three N$_2$H$^+$ maps obtained from a MOPRA survey \citep{jones2012}. The images are integrated over different velocity ranges in order to emphasise coherent structures. Regions identified in the X-ray map are overlaid: they appear in three different velocity ranges. The molecular cloud associated with the Arches cluster appears in a fourth range around $-25$~km~s$^{-1}$, which is not shown here.}
	\label{fig:MOPRA}
\end{figure*}
The bright regions of the Sgr~A complex (MC1, MC2, Br1, Br2 and G0.11-0.11) and the Arches cluster are identified by ellipses in Figure~\ref{fig:ChandraView} (left panel). The cloud designation is similar to the one used by \citet{ponti2010}, except that the Bridge region is now split in two clouds: Br1 and Br2.
In addition to these bright regions, a lower level of emission is also visible, having the shape of two large filaments going north of MC1 towards the Arches cluster, and the hint of a third and even fainter structure extending south of MC2 and Br1.
From this map it is clear that a significant fraction of the region has been illuminated for at least part of the past decade.\\\\
An \fe\ map showing the distribution of illuminated features seen in July 2011 by Chandra is also presented in Figure~\ref{fig:ChandraView} (right panel). From this map we see that the brightest emission was then in the eastern part of MC1 and in the Bridge, with in particular a new elongated bright structure in the center of Br2. Almost no bright emission is left in the other two areas.\\\\
Following the method of \citet{ponti2010} we used molecular line data to identify the X-ray emitting regions that are correlated with coherent molecular structures. The N$_2$H$^+$ molecule is a good tracer for dense dark clouds and happens to best match the distribution of X-ray emission seen in the Sgr~A complex. Therefore, we used the N$_2$H$^+$ $J$=1-0 data-cube provided by the MOPRA CMZ surveys \citep{jones2012} with an angular resolution around 40~arcsec. 
%\\\\%
The three maps shown in Figure~\ref{fig:MOPRA} are obtained by integrating the N$_2$H$^+$ signal over different velocity ranges. Sgr~A regions identified in the Chandra \fe\ map are clearly visible in MOPRA maps with distinct velocity ranges. The MC1 and MC2 clouds are visible around $-10$~km~s$^{-1}$. For these two clouds the N$_2$H$^+$ map and the 1999--2011 X-ray mosaic (Figure \ref{fig:ChandraView}) show very similar features with in particular a dense blob at the bottom east of MC2 and linked to MC1 by a thin feature. The G0.11-0.11 cloud is visible around $+20$~km~s$^{-1}$ and the N$_2$H$^+$ emission highlights the two elongated features of this structure that are also visible in the 1999--2011 X-ray mosaic. The two parts of the Bridge region, Br1 and Br2, are visible around $+50$~km~s$^{-1}$. From the molecular line data we can infer that there are three coherent and distinct structures. Their respective column densities and positions along the line of sight cannot be derived from these maps and will be discussed along with the different possible interpretations of the X-ray reflection in these structures (section \ref{sec:conclu}).
\subsection{Time variations of few arcmin scale regions and comparison with previous works} \label{sec:FeKacloud}
\begin{figure*}
	\centering
	\includegraphics[trim = 0mm 0mm 0mm 0mm,clip,width=1\textwidth]{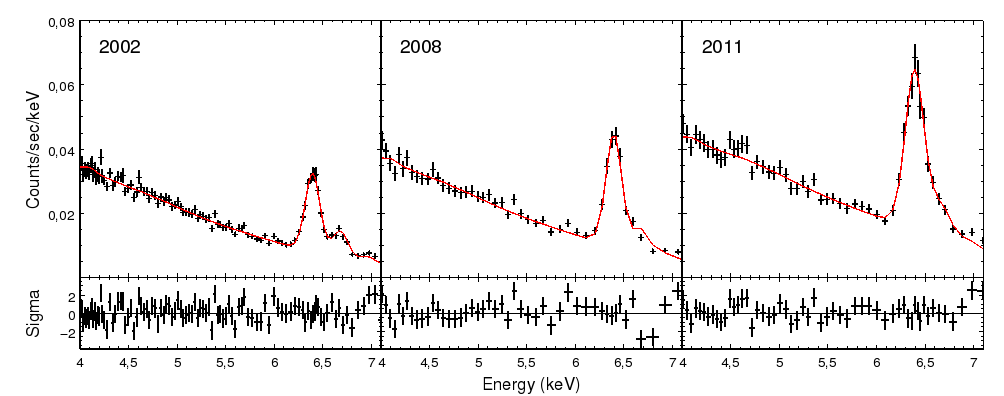}
	\caption{Bridge spectra and fit as in 2002 (no bright illumination seen in the Bridge), 2008 (illumination starts in the western part of Br1) and 2011 (illumination reaches Br2). The \fe\ flux is significantly increasing over the years.}
	\label{fig:Bspectra}
\end{figure*}

The neutral iron~K$\alpha$ line emission from molecular clouds is strongly variable in the inner regions of the Galaxy, providing the evidence that an important part of the diffuse emission is due to reflection. Therefore, studying these variations is crucial to constrain the illuminating event.
The general trend of these variations has already been characterized on arcminute scale structures \citep{ponti2010, capelli2012} and we perform here a similar analysis of the Chandra data.\\\\
Samples of spectra and corresponding fits obtained on the Bridge region \citep[as defined by][and including Br1 and Br2]{ponti2010} are presented in Figure \ref{fig:Bspectra}. These three plots show three relevant periods of the progressive illumination of the Bridge structure (2002, 2008 and 2011).
The nine overall Bridge spectra were fitted simultaneously as explained in section \ref{sec:DA-specfit}. The fit results specific to the three spectra plotted in Figure \ref{fig:Bspectra} are presented in Table~\ref{tab:specBridge}.
\begin{table}[h]
	\centering
	\caption{2002, 2008 and 2011 best fit parameters for the overall Bridge region. 
	}
	\begin{scriptsize}	
	\begin{tabular*}{0.5\textwidth}{@{\extracolsep{\fill}} l c c c}
	\hline \hline
	Parameters & 2002 & 2008 & 2011 \\
	\hline 
\fe\ Norm ($10^{-5}$~ph~cm$^{-2}$~s$^{-1}$) & $3.8 \pm 0.1$ & $6.0 \pm 0.3$ & $8.5^{+0.2}_{-0.3}$\\ [1.2ex]
C. Photon Index & $0.5\pm 0.1$ & $0.2 \pm 0.1$ & $0.0 \pm 0.1$\\ [1.2ex]
C. Norm  ($10^{-5}$ ph~kev$^{-1}$~cm$^{-2}$~s$^{-1}$) & $3.3 \pm 0.1$ & $4.3 \pm 0.2$ & $5.6 \pm 0.2$\\
	\hline \hline
	\end{tabular*}
	\label{tab:specBridge}
		\tablefoot{
The nine year spectra were fitted simultaneously ($\chi^2_{gehrels}~/~\text{d.o.f.}~=~1148~/~1395$) giving for each year the \fe\ line flux, the continuum photon index and its normalization. We point out once more that the continuum emission is modelled by a simple power-law and that the physical interpretation of its parameters is not straightforward. The hot plasma normalization obtained for the Bridge region fit is $(1.83\pm0.08)\times10^{-3}$~cm$^{-5}$. The other parameters were fixed following section \ref{sec:DA-specfit}.
	}
	\end{scriptsize}	
	\vspace{-0.4cm}
\end{table}\\\\
On corresponding years, the fit results are fully consistent with the previous analysis, for all regions. Our new spectral characterization also provides three more data points (2000, 2010 and 2011) to the \fe\ flux lightcurve of each region and thereby increases the significance of the variations.
On a large scale, the Bridge \fe\ flux has significantly increased (linear regression $20\,\sigma$ better than constant fit) and we report the illumination of the Br2 region in 2011, confirming the propagation along the Bridge predicted by \citet{ponti2010}.
In the MC2 cloud, the \fe\ flux has clearly decreased ($6.8\,\sigma$), which confirms the trend suggested by \citet{capelli2012}.
The emission of the overall MC1 cloud has not shown significant variations ($2.7\,\sigma$ and down to $1.8\,\sigma$ if we remove the 2000 data point which is $3\,\sigma$ lower than the constant fit).
Concerning G0.11-0.11, the coverage of the entire region is not sufficient to draw conclusions about its overall emission variation, but its generally decreasing trend will be characterized by a smaller scale analysis in section \ref{sec:48small}. \\\\
Therefore, our arcminute scale characterization of the variations confirms the general trends described in previous works. 
\begin{figure*}
	\centering
	\includegraphics[trim = 7mm 8mm 7mm 16mm,clip,width=0.75\textwidth]{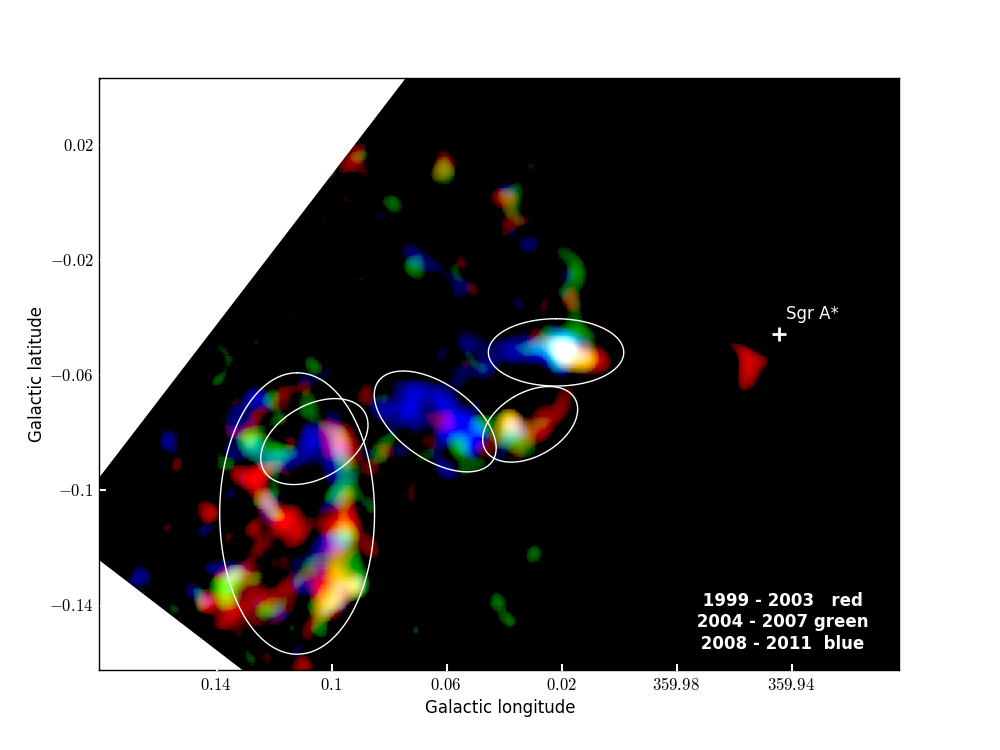}
	\caption{\fe\ flux mosaics smoothed to 9~arcsec. \textit{(Red)} Chandra data from 1999 to 2003 (only observation 3392 has been kept among 2002 observations); \textit{(Green)} Mosaic map from 2004 to 2007; \textit{(Blue)} Mosaic map from 2008 to 2011. The emission has been strongly varying in the Sgr~A complex with a clear trend from west to east in MC1 and MC2, a late illumination of Br1 and Br2, and more complex variations in G0.11-0.11. We assume Br2 emission is only associated with the blue patch at the center of the region, the rest being mostly due to the underlying G0.11-0.11 cloud.}
	\label{fig:RGB}
\end{figure*}
\subsection{Motivations for smaller scale characterization}
Our previous analysis of the arcminute scale structures assumes the X-ray emission is due to coherent and independent clouds. Nevertheless, this is often not the case since different structures can be positioned along the line of sight, as it is the case for G0.11-0.11 and the Br2 structures. Moreover such an analysis is unable to characterize variations smaller than the cloud physical size (about 8 light-years) and is therefore unable to fully characterize the accurate profile of \sgr's past emission in case of fast variations. \\\\
In order to visualize the emission variations inside these large structures we build a RGB map with the three colors corresponding to three different time periods between 1999 and 2011. 
This map, presented in Figure~\ref{fig:RGB}, shows that all emitting areas have been varying, including the MC1 cloud. This last structure, characterized as constant by the larger scale analysis, is in fact composed of subregions that have been varying differently (see section \ref{sec:MC1} for further characterization). Moreover, variations are not random. The MC1 emission has been moving from west before 2003 to east after 2008. In MC2, the signal has been going from west before 2003 to mainly off after 2008, and the Bridge seems to be a coherent area becoming illuminated between 2008 and 2011. G0.11-0.11 variations are more complex as illuminated patches are appearing and disappearing in all three periods.
\\\\
A spectral analysis performed on the full cloud scale hides small scale variations highlighted instead by both image comparison and spectral analysis on smaller scales (section \ref{sec:FeKasmall}). Therefore, working on large scales in order to obtain high-statistic spectra to fully characterize the emission can be double-edged. Indeed, large-scale spectra are de facto including different structures that are not equally illuminated. This is why large-scale spectral parameters such as the iron line intensity and its equivalent width have to be interpreted with great caution. For this reason we decided not to work any further on large-scale spectra but to use Chandra high angular resolution in order to further characterize variations on the smallest possible scale.
%
%
%
%
%
%%%%%%%%%%%%%%%%%%%%%%%%%%%%%%%%%%%% FeKa small regions
\section{\fe\ emission: variations on a scale of $26''\times61''$}
\label{sec:FeKasmall}
This section focusses on the analysis of two bright regions: the Bridge and MC1. These molecular clouds show strong \fe\ variations but with two different behaviors.
%
%%%%%%%%%%%%%%%%% Bridge Variations
\subsection{The Bridge: strong variation on a short (year) time scale}\label{sec:Bridge} 
\begin{figure}
	\centering
	\includegraphics[trim = 0mm 0mm 0mm 0mm,clip,width=0.5\textwidth]{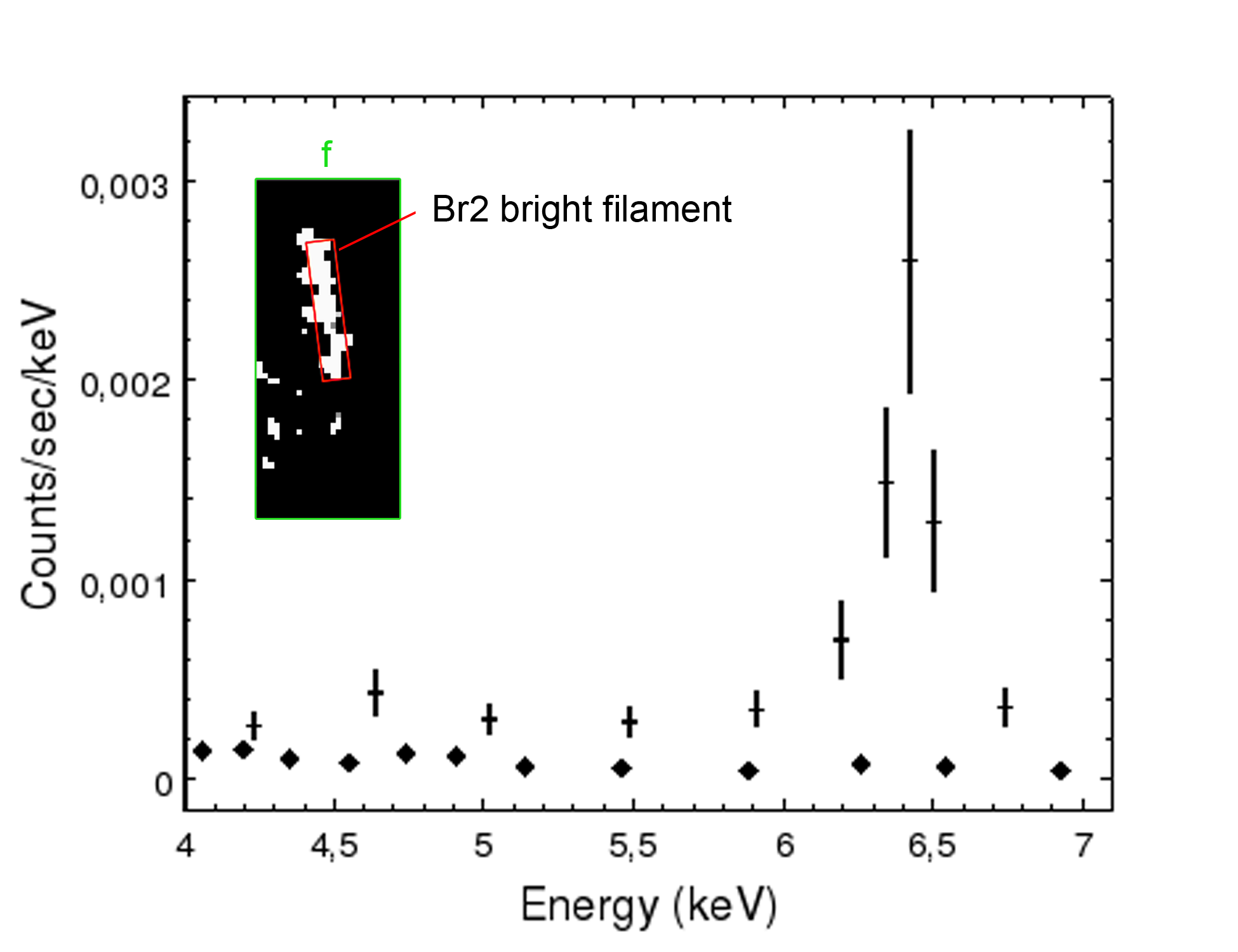}
	\caption{The 2011 Br2 bright filament. The \fe\ flux map in the upper left corner is a zoom on region f which is defined in Figure \ref{fig:Bridge}. Here the map has not been smoothed and shows the exact dimension of the bright illumination detected in 2011. Both spectra presented on this figure are extracted from the region highlighted by the red rectangle. The 2011 spectrum shows an \fe\ emission line at 6.4~keV (crosses), while the 2000 to 2010 spectrum (diamonds) shows no excess at this energy. Notice that due to better statistics the error bars of the \hbox{2000--2010} spectrum are not larger than the diamond size. The 2011 spectrum subtracted from the 2000--2010 one is fitted over the 4--7.1~keV energy range ($\chi^2_{gehrels} / \text{d.o.f.} = 7.8 / 6$): the \fe\ line intensity is $(3.2\pm0.6)\times10^{-6}$~ph~cm$^{-2}$~s$^{-1}$ and the equivalent width is more than 2~keV, suggesting a higher metallicity than previously anticipated.
	}
	\label{fig:Br2spec}
\end{figure}

\begin{figure*}
	\centering
 	 \includegraphics[trim = 0mm 0mm 0mm 0mm,clip,width=0.9\textwidth]{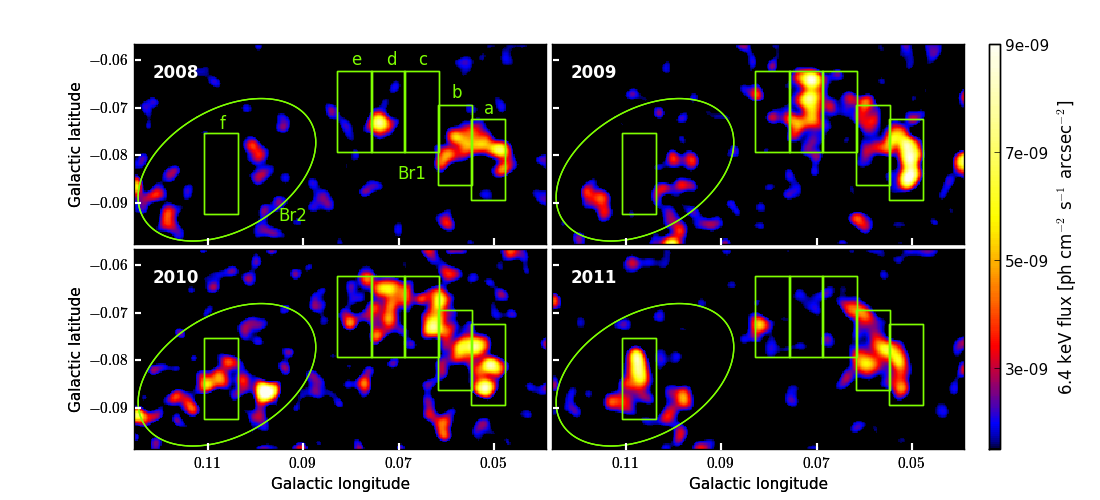}
	\caption{\fe\ flux mosaics showing the evolution of the Bridge region from 2008 to 2011. Br1 is divided into five subregions named from a (west) to e (east). This region becomes brighter from 2008 to 2010. In 2011, the eastern part of Br1 is mainly off and the Br2 region gets very bright (box f). The exact illuminated region of Br2 is shown in Figure \ref{fig:Br2spec} and appears larger here because it is smoothed to 4~arcsec. 
	}
	\label{fig:Bridge}
\end{figure*}

The previous RGB map gives a clear view of the variations occurring in the Sgr~A complex region except maybe for the Bridge area that has been varying only in the last few years. Figure~\ref{fig:Bridge} gives a more precise view of the evolution of the Bridge emission during the last four years. In particular, it confirms that Br2 was not strongly illuminated before 2011. Its 2011 emission is then relatively strong and condensed into a very thin region having physical dimensions at 8 kpc of about $0.2\times1$~pc$^2$ and it is located at least 23.4~pc from \sgr. We extract the 2011 spectrum of the bright region highlighted by a red rectangle in Figure~\ref{fig:Br2spec} and we take the sum of all previous year spectra as background spectrum.  
Both are shown in Figure \ref{fig:Br2spec}.
\\\\
The subtracted spectrum only includes the variations induced by the reflection and can be simply fitted by a power-law continuum and a gaussian iron K$\alpha$ line emission.
The \fe\ line intensity best fit is $(3.2\pm0.6)\times10^{-6}$~ph~cm$^{-2}$~s$^{-1}$ ($\chi^2_{gehrels} / \text{d.o.f.} = 7.8 / 6$). 
By considering the illuminated region has a filamentary shape, we assume that its depth along the line of sight is no more than its width, i.e., about 0.2~pc. We assume that its column density is around $10^{23}$~cm$^{-2}$ \citep[consistent with the overall Bridge column density derived by][]{ponti2010}, it would mean that the filament density is around $2\times10^{5}$~cm$^{-3}$, which is already conservative. 
Following \citet{sunyaev1998}, we can then compute the minimum \sgr\ luminosity needed to illuminate this region. We find that the luminosity between 2 and 10~keV had to be at least of the order of 10$^{39}$~erg~s$^{-1}$. The event illuminating this region is therefore a strong one. Moreover, the region brightens in less than one year, which implies that the luminosity of the illuminating source increased suddenly, by at least a factor~10. We point out once more that a lower-resolution analysis cannot characterize such events.
\begin{figure*}
	\centering
	\includegraphics[trim = 35mm 0mm 35mm 2mm,clip,width=1\textwidth]{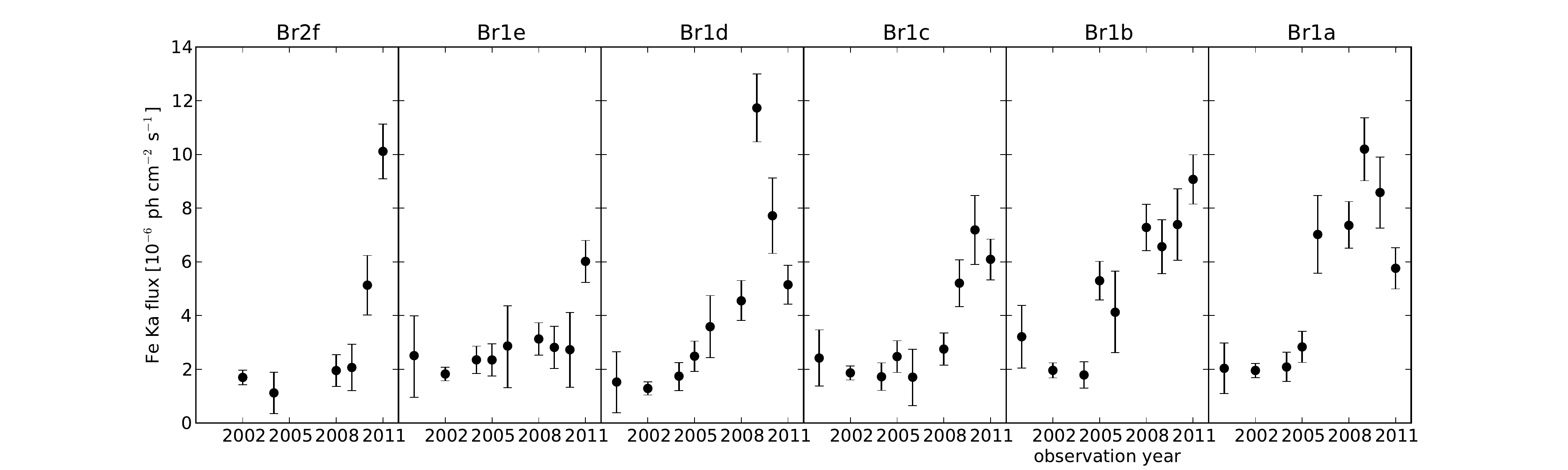}
	\caption{\fe\ flux lightcurves corresponding to six $26\times61$~arcsec$^2$ rectangular regions covering the bright emission of the Br1 and Br2 clouds. Subregions are named from f (east) to a (west). 
	All lightcurves show a sharp increase shifted in time from region to region. }
	\label{fig:Bsubspec}
\end{figure*}\\\\
The illumination of Br2 supports the propagation along the Bridge predicted by \citet{ponti2010}. However, the proposed scenario did not anticipate the high source variability implied by the fast variations revealed by Chandra in all regions, including the Bridge. 
The comparison of Br1 images between 2008 and 2011 shown in Figure~\ref{fig:Bridge} reveals a fast illumination of the area from west to east until 2010 and then an abrupt fading of Br1 eastern part in 2011 (see Br1d lightcurve in Figure \ref{fig:Bsubspec}). 
We perform a spectral analysis in order to prove that the variations seen in the image are significant. \fe\ fluxes are extracted from the $26\times61$~arcsec$^2$ rectangles shown in Figure~\ref{fig:Bridge} and the corresponding lightcurves are presented in Figure \ref{fig:Bsubspec}. The bright structure around (l,b)$\,=\,$($0.100^\circ$,~$-0.085^\circ$) is ignored because it undergoes complex variations of both shape and flux which are difficult to reconcile with either Br2 or the underlying G0.11-0.11 cloud variations.\\\\
All lightcurves show significant variations as the increasing linear regression is preferred over the constant fit with at least $4.5\,\sigma$ and up to $10\,\sigma$ confidence level. 
In addition, a simple linear fit does not represent well the data, whose variations are characterized by a sharp increase shifted in time from region to region. This pattern strengthens the scenario of a propagation along the Bridge structure \citep{ponti2010} and indicates that the 2011 bright filament is indeed part of the Br2 cloud rather than G0.11-0.11. 
The 2011 filament is included in the Br2f region whose lightcurve is compatible with the previous analysis performed on the filament alone. This larger scale analysis is less relevant to precisely characterize the variation but has the advantage of being directly comparable with Br1 subregion variations.
The fastest variation is seen in the Br1d region, which peaks at $(11.7\pm1.3)\times10^{-6}$~ph~cm$^{-2}$~s$^{-1}$ with about a factor 10 increase compared to the 2002 flux value. The time scale of the variation is less than three years, since the \fe\ flux in both 2008 and 2011 is down to less than half of its peak value.
This trend seems to be due to the illumination of an isolated filament as in Br2f and is compatible with the variation seen in region Br1a if we exclude the 2006 data point (mainly due to a bright structure that is no longer visible after then). Region Br1b might include more than one clump and thus misses the short variation.
If the flare illuminating the Bridge region is 2~year-long, we expect Br1c and Br2f regions to drop to zero emission in the next few years while Br1e region should increase again before dropping as the others.
\\\\
We conclude that the Br1 and Br2 regions are likely to be illuminated by a single strong event lasting less than three years. Moreover, the position of the western filament in the Br1a subregion seems to be moving away from \sgr\ at a velocity compatible with the speed of light. This is a hint that the illuminating event could be even shorter but the extraction boxes are too large to fully characterize it.
%
%%%%%%%%%%%%%%%%%% MC1 variations
\subsection{MC1: linear variations on longer (decade) time scale}\label{sec:MC1}
\begin{figure*}
	\centering
	\includegraphics[trim = 0mm 0mm 15mm 2mm,clip,width=1\textwidth]{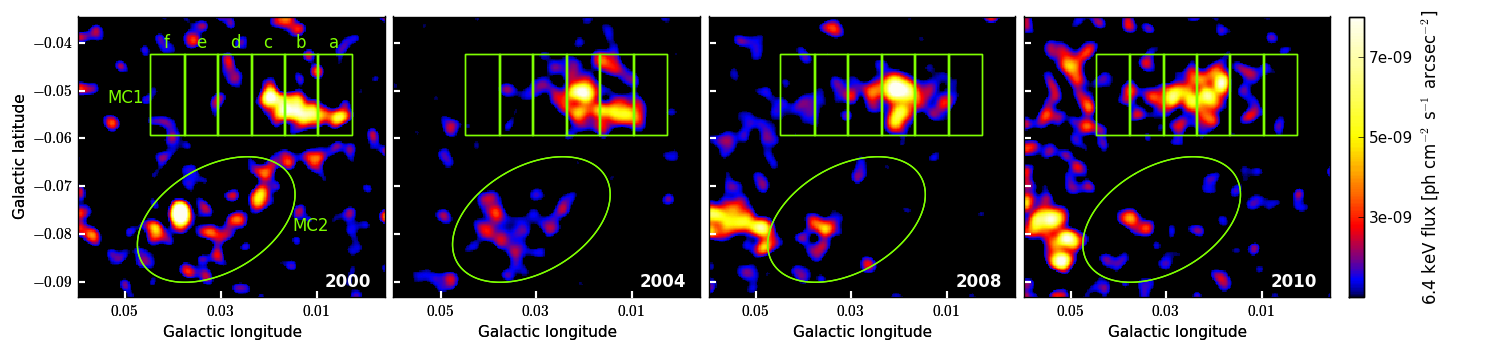}
	\caption{\fe\ flux maps showing the signal propagation along the MC1 and MC2 structures from 2000 to 2010. The six subregions of MC1 are those used for the spectral analysis. The propagation (away from the Galactic center) is visible in both clouds. The 2010 image looks clumpier due to lower exposure time.}
	\label{fig:MC1prop}
\end{figure*}
\begin{figure*}
	\centering
	\includegraphics[trim = 35mm 0mm 35mm 2mm,clip,width=1\textwidth]{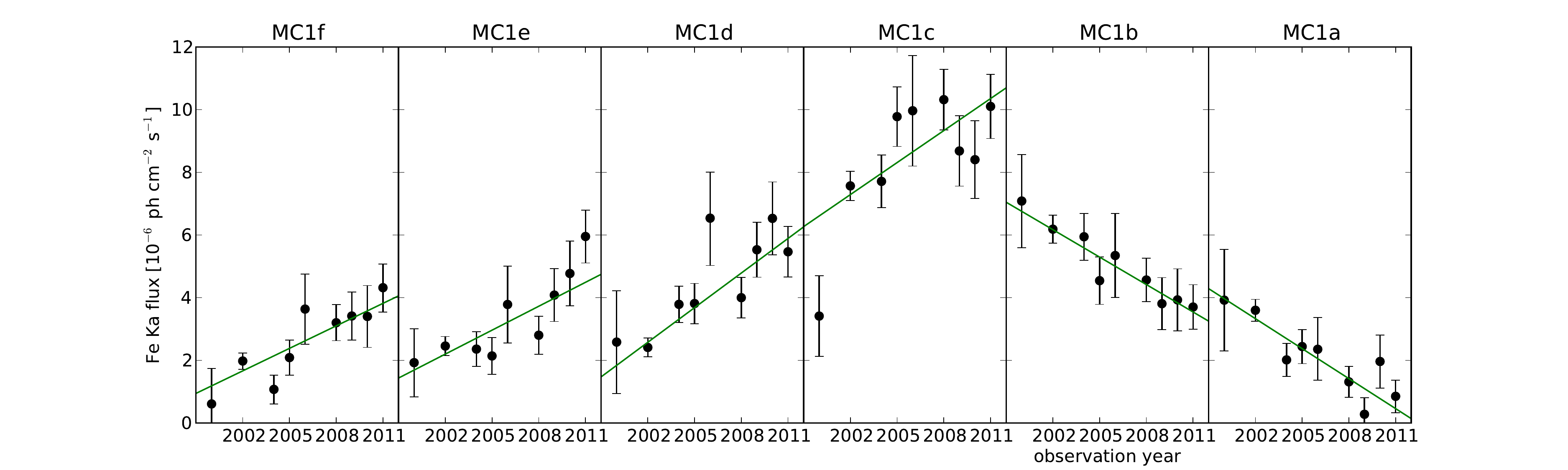}
	\caption{\fe\ flux lightcurves corresponding to six $26\times61$~arcsec$^2$ rectangular regions covering the bright emission of the MC1 cloud. Subregions are named from f (east) to a (west). Linear regression lines (in green) are in good agreement with the data and have similar absolute slopes. }
	\label{fig:MC1lc}
\end{figure*}
The MC1 cloud emission at 6.4~keV was found constant by the previous arcminute-scale analyses \citep{ponti2010,capelli2012}. We confirm that on arcminute scale the \fe\ emission of this molecular cloud is compatible with a constant emission. Nevertheless, the unique spatial resolution of Chandra allows us to study the variations of the \fe\ morphology on arcsecond scale.  Both the RGB map in Figure \ref{fig:RGB} and the sample of enlargements presented in Figure \ref{fig:MC1prop}, show that during the past decade the emission has varied with a clear pattern (from west to east). 
Therefore, smaller regions show faster variations.
In order to characterize MC1 variations, we choose the subdivision presented in Figure~\ref{fig:MC1prop}. From the 2000, 2004, 2008 and 2010 images of MC1 it is already clear that the subdivision chosen is still large to fully characterize the signal variations since the illuminated regions are small compared to the box size. Nevertheless, spectral analysis on even smaller regions does not give significant results due to poor statistics.\\\\
The lightcurves built thanks to the \fe\ line fit performed on MC1 subregions are shown in Figure~\ref{fig:MC1lc}. All lightcurves are characterized by a linear regression which is preferred with a confidence of $3.9$ to $5.8\,\sigma$ over a constant fit. The two western subregions (MC1a, b) display a decrease and the three eastern ones (MC1d, e, f) an increase of the emission level. 
The MC1c lightcurve is not well fitted by the linear regression. 
Its 2000 \fe\ flux stands $3.8\,\sigma$ below the constant fit and if we remove it from the fit the significance of the linear regression over the constant fit drops to only $2.8\,\sigma$. This is a hint that MC1c emission has reached a maximum luminosity and extrapolating from the emission of its nearby structures, we suggest that the MC1c region will experience the same decrease as MC1a and MC1b in the next few years.\\\\
Therefore, the MC1 variations are consistent with being the result of a signal propagating through a molecular cloud. Moreover, the absolute slopes of the subregion regressions are all consistent with a temporal gradient of $0.3\times10^{-6}$~ph~cm$^{-2}$~s$^{-1}$~yr$^{-1}$ which is consistent with a rather homogeneous and continuous structure.
\\\\
This is the first reported detection of 6.4~keV variations inside the MC1 cloud. These variations closely follow a linear time variation, decreasing close to the Galactic center and increasing farther away, and appear quite different from what is seen in the Bridge region. These differences can be due to either distinct cloud structures or distinct illuminating events. These two possibilities will be discussed in section \ref{sec:conclu}.
%
%
%
%
%%%%%%%%%%%%%%%%%%%%%%%%%%%%%%%%%%%% GLOBAL VARIATIONS
\section{The 4--8 keV emission: systematic analysis of the Sgr~A complex variations on a $15''$ scale}
\label{sec:48small}
The \fe\ characterization presented in \textbf{section \ref{sec:FeKasmall}} was performed on specific and bright regions where such an analysis was feasible. It highlights the importance of small-scale analysis but cannot be performed on the overall Sgr~A complex since the statistics are too poor in most regions. Therefore, we have performed a systematic analysis of the overall Sgr~A complex region using the 4 to 8~keV flux to characterize the high-energy signal variations in 15~arcsec squared bins. The aim is to determine the time scales of all varying regions in the complex in order to constrain the illumination pattern as much as possible.\\\\
Since the instrumental background is weak compared to the typical emission in the Sgr~A complex, and since the only varying extended component is the reflection, we can safely attribute the variations to the propagation of energetic X-rays. Indeed, the test performed on the previously studied regions shows that the 6.4~keV and the 4--8~keV flux emissions display the exact same variations, which thereby indicates that the 4--8~keV flux is a proper tracer of the variations due to reflection. It allows an improved statistical characterization of the variations but the origin of its steady component is of course more difficult to assess, since there are contributions from the hot thermal plasma and the internal background.
\subsection{Presentation of the data through images}
We performed a systematic analysis of the data-cube containing the 4--8~keV flux information for each 15~arcsec squared pixel of the Sgr~A complex map and for each available year.
We use the CIAO \textit{dmstat} routine to count the number of photons of energy between 4 and 8~keV  in the image and in the background image. Using the corresponding exposures we then compute the 4--8~keV flux for each year and the corresponding $1\,\sigma$ error, assuming a normal distribution.
\begin{figure*}
	\centering
	\includegraphics[trim = 2mm 8mm 5mm 18mm,clip,width=0.95\textwidth]{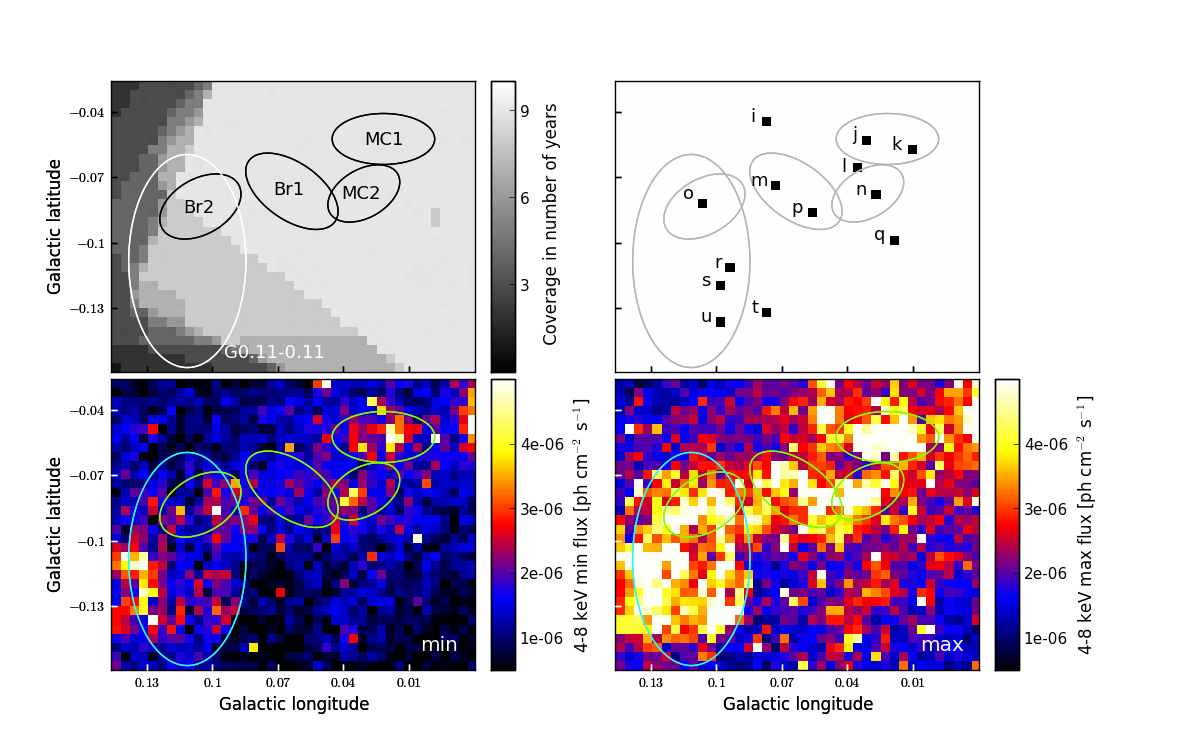}
	\caption{ Information maps presenting the data-cube used in the 4 to 8~keV flux characterization. The pixels correspond to 15~arcsec squared regions from which the lightcurves were extracted. For each region we plot: \textit{(Top left)} the number of lightcurve data points; (\textit{Bottom left}) the lightcurve minimum value; \textit{(Bottom right)} the lightcurve maximum value. \textit{(Top right)} Positions of individual lightcurves presented in Figures \ref{fig:null}, \ref{fig:Br12lc}, \ref{fig:MC12lc} and \ref{fig:G011lc}.}
	\label{fig:15data}
\end{figure*}\\\\
Due to different observation strategies and the limited field of view, the time coverage is not uniform across the studied regions. The MC1, MC2, Br1 and Br2 areas are well covered (data from 8 or 9~years) while G0.11-0.11 presents a sparser coverage on the western and southern parts (only 2 or 3~years). 
The number of observation years available for each pixel is shown on the top left map in Figure \ref{fig:15data}, and it corresponds to the number of data points constituting the corresponding flux lightcurves. The bottom left and right panels of Figure \ref{fig:15data} present respectively the minimum and the maximum values of all pixel lightcurves.
Both maps show that the emission is correlated with molecular material, as denser regions appear brighter. Minimum values are not representative of a zero level of emission since part of the map region is illuminated during the entire time period. 
Yet, the minimum flux values are almost all below $2.5\times10^{-6}$~ph~cm$^{-2}$~s$^{-1}$ while the maximum map highlights the five molecular structures previously identified, the cores of which are detected at more than $5\times10^{-6}$~ph~cm$^{-2}$~s$^{-1}$. From this comparison it is clear that variations are visible on a 15 arcsec scale and they might not be limited to the brightest regions. In order to further characterize these variations, we perform a systematic analysis of the lightcurve variations. Some characteristic lightcurves of regions pointed out on the top right panel in Figure \ref{fig:15data} will be presented in Figures \ref{fig:null}, \ref{fig:Br12lc}, \ref{fig:MC12lc} and \ref{fig:G011lc} to further illustrate the different variation behaviours.
\subsection{Characterization of two variation behaviors}
\begin{figure*}
	\centering
	\includegraphics[trim = 2mm 8mm 5mm 18mm,clip,width=0.95\textwidth]{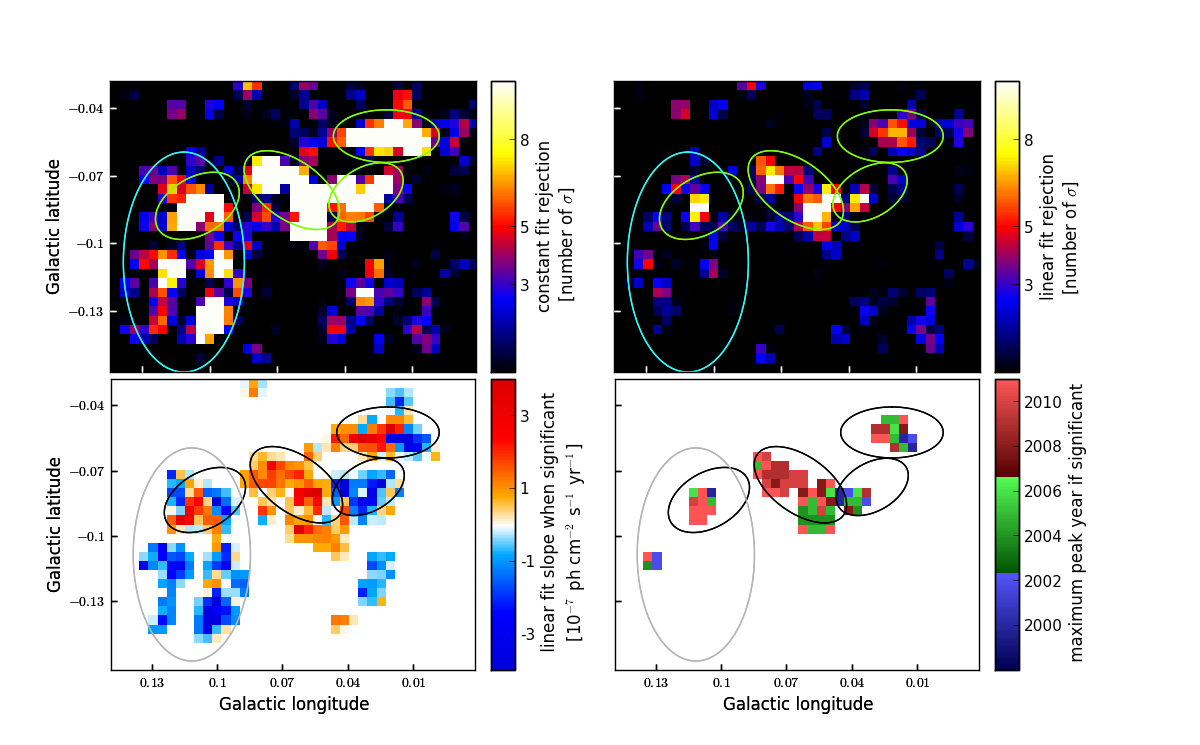}
	\caption{ Information maps derived from the 4 to 8~keV flux lightcurve analysis. The main point sources have been removed before the analysis. Probabilities displayed in the two top maps are combined to a scale of 30~arcsec and represent: \textit{(Top left)} post trial rejection of a constant fit in number of sigma; \textit{(Top right)} post trial rejection of a linear fit in number of sigma. In the two bottom maps are plotted: \textit{(Bottom left)} the linear fit slope of the lightcurve if the constant fit is rejected at more than $5\,\sigma$, zero otherwise; \textit{(Bottom right)} the year corresponding to the lightcurve's maximum value when the linear fit is rejected at more than $5\,\sigma$. From these figures we see that variations are correlated between nearby regions and are either linear (MC1, MC2 and G0.11-0.11) or not (Br1 and Br2). Moreover, both linear fit slope and peak year characterizations highlight an apparent signal propagation from right (decreasing) to left (increasing).}
	\label{fig:6variations}
\end{figure*}

\begin{figure*}
	\centering
	\includegraphics[trim = 0mm 0mm 0mm 0mm,clip,width=0.65\textwidth]{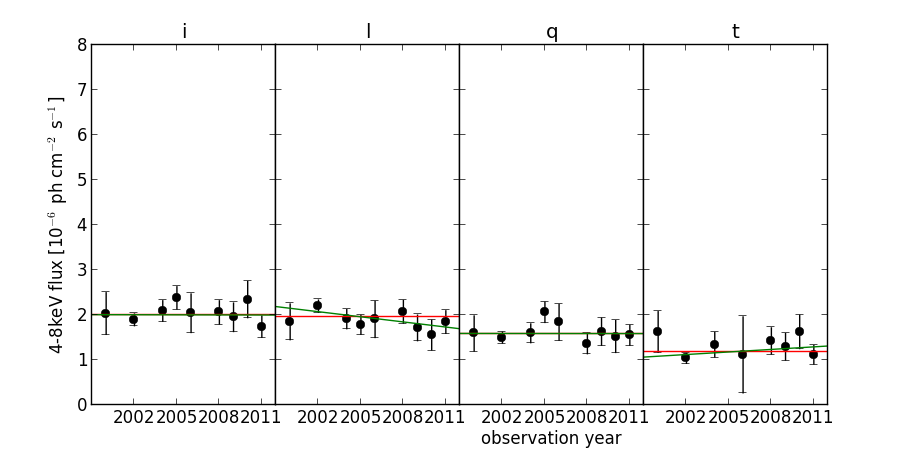}
	\caption{The 4 to 8~keV lightcurves of different 15-arcsec-square regions overlaid with constant (red) and linear (green) fits. The exact positions of these regions are given in Figure \ref{fig:15data}. Their emissions are all described as non-varying by the previous analysis and their lightcurves are compatible with constant intensity.}
	\label{fig:null}
\end{figure*}

\begin{figure}[h]
	\centering
	\includegraphics[trim = 0mm 0mm 0mm 0mm,clip,width=0.5\textwidth]{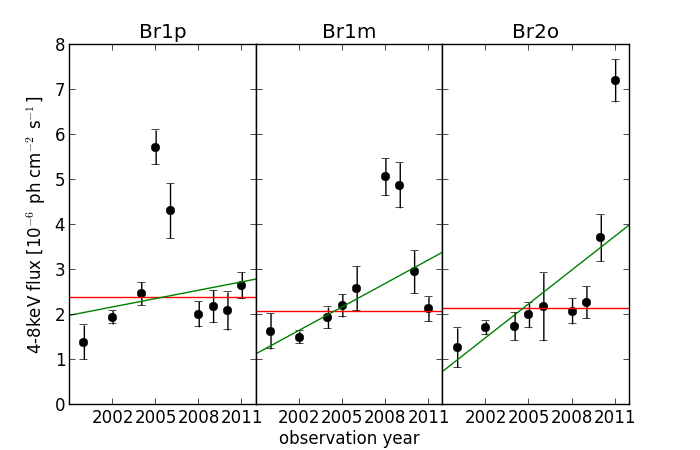}
	\caption{The 4 to 8~keV lightcurves of different 15-arcsec-square regions in Br1 or Br2, overlaid with constant (red) and linear (green) fits. The exact positions of these regions are given in Figure \ref{fig:15data}. The fast variations seen in these regions are poorly fitted by the linear regression.}
	\label{fig:Br12lc}
\end{figure}

\begin{figure}[h]
	\centering
	\includegraphics[trim = 0mm 0mm 0mm 0mm,clip,width=0.5\textwidth]{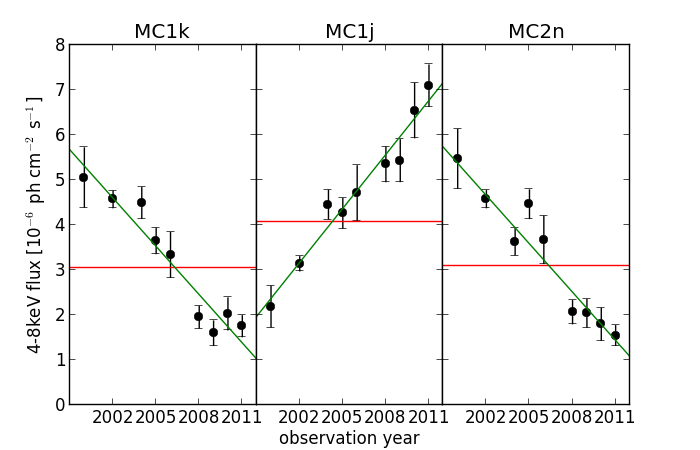}
	\caption{The 4 to 8~keV lightcurves of different 15-arcsec-square regions in MC1 and MC2, overlaid with constant (red) and linear (green) fits. The exact positions of these regions are given in Figure \ref{fig:15data}. The strong variations seen in these regions are well fitted by a linear regression.}
	\label{fig:MC12lc}
\end{figure}

\begin{figure}[h]
	\centering
	\includegraphics[trim = 0mm 0mm 0mm 0mm,clip,width=0.5\textwidth]{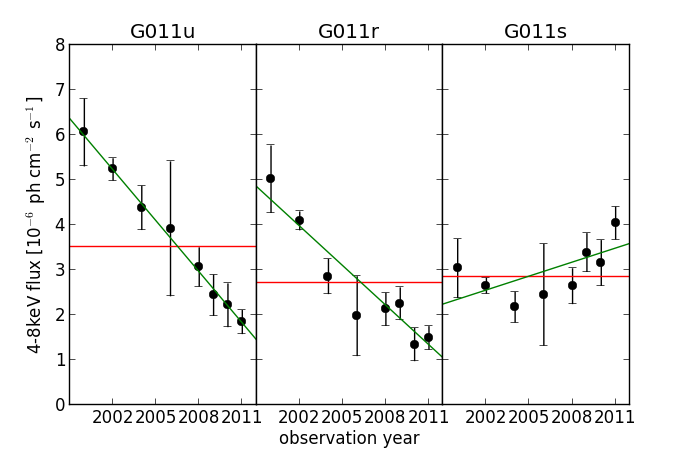}
	\caption{The 4 to 8~keV lightcurves of different 15-arcsec-square regions in G0.11-0.11, overlaid to constant (red) and linear (green) fits. The exact positions of these regions are given in Figure \ref{fig:15data}. The strong variations seen in the first two regions are well fitted by a linear regression, and the third hints for a different behavior.}
	\label{fig:G011lc}
\end{figure}

To characterize the regional variations we perform both linear and constant least-squares fits to the 4--8~keV lightcurves extracted from 15~arcsec squared regions. 
The results are presented in terms of fit rejection probabilities in the two top maps of Figure \ref{fig:6variations}. Regions presenting a significant point-source contribution have been removed for this analysis. Moreover, in order to improve the visibility of regions displaying similar behaviors, probabilities are combined on a scale of 30~arcsec. We take the number of trials into account and correct probabilities accordingly.
\\\\
The constant fit results highlight the regions that are significantly varying (top left panel in Figure~\ref{fig:6variations}). They are mostly included in the five large identified regions and represent about one third of the total area. About two thirds of the Sgr~A complex is therefore characterized as constant by our analysis. These subregion's lightcurves are compatible with a constant emission, as shown in Figure~\ref{fig:null}. Their fluctuations are negligible and this is strong evidence that variations detected in the regions that correlate with the molecular structures are real.
\\\\
The linear fit results highlight the regions having a strong non-linear variation (top right panel in Figure~\ref{fig:6variations}). They are found mostly in Br1 and Br2 and a sample of lightcurves extracted from these regions is presented in Figure \ref{fig:Br12lc}. The time scale of the variation observed is fully compatible with what has been inferred from the previous \fe\ characterization of these regions, and the amplitude of the variations is slightly smaller because in the 4--8~keV range the reflection emission is polluted by non-varying background emissions.
\\\\
By comparing the linear and the constant fit results we also identify regions which are characterized by a linear variation. Such regions are mainly included in the MC1, MC2 and G0.11-0.11 structures. A sample of the lightcurves from MC1 regions is shown in Figure \ref{fig:MC12lc}. They are fully compatible with what has been deduced from the spectral analysis of this region and the variation amplitude obtained by this second analysis is even slightly larger because the size of the subregions we consider here are closer to the emission variation angular size. Moreover,  MC2 and G0.11-0.11 cloud emissions are globally decreasing on a ten-year time scale and a sample of characteristic lightcurves of these two regions is presented in Figures \ref{fig:MC12lc} and \ref{fig:G011lc}. The emission decrease is fully compatible with the one seen in the MC1 cloud and this is a strong hint that these three clouds are witnessing the same illuminating event. Nevertheless, some subregions of G0.11-0.11 seem to have a different trend with, in particular, a hint of a slight emission increase. This difference inside G0.11-0.11 can be explained by a complex structure of the molecular cloud as these particular subregions could either be slightly further away along the line of sight and therefore seeing the same event but with a delay, or they could be slightly closer and seeing a later event.

\subsection{Hints of a signal propagation}
From the previous individual lightcurve characterizations it is clear that most of the variable regions have linear variations. In order to visualize the spatial disposition of the increasing and decreasing trends, we map the value of the linear fit slope if the constant fit is rejected at more than $5\,\sigma$. The results are shown on the bottom left panel of Figure~\ref{fig:6variations}.
Most detected variations are spread over several contiguous regions showing the same trend. In particular, all five identified areas show correlated variations. The signal appears to have propagated from west to east as the western part of MC1 and MC2 have shown a decrease while the eastern part of MC1 and the Bridge have increased. In the easternmost portion of the field, the G0.11-0.11 cloud does not conform to this apparent propagation; it shows a clear decrease. However, if located much further in the foreground, the G0.11-0.11 cloud could be seeing the same X-ray event as the regions showing the West to East trend. This method of mapping the slope of the variability also highlights variations in fainter areas by detecting a decrease to the south of the MC2 area and an increase to the south of the Br1 area. This strengthens the theory that an X-ray front has been propagating through these structures and that the fainter areas are spatially nearby the brighter areas. 
\\\\
The linear fit is inadequate to properly characterize Br1-type variations. Therefore, we also consider the year of the peak emission if the linear fit is rejected at more than $5\,\sigma$. The results are mapped in the bottom right panel of Figure~\ref{fig:6variations}. According to this stricter characterization, 7\%  of the total area has non-linear variations.
Most of the variations are detected within the Bridge and the peak years are fully consistent with the previous linear characterization. We point out that the MC1 central region also presents non-linear variations, as previously mentioned in section \ref{sec:MC1}. This non-linear trend is restricted to a few central pixels but is spread over a larger region of the map due to the smoothing of the probabilities over a 30~arcsec scale.
\\\\
The systematic characterization of the variations in the Sgr A complex confirms the presence of two different time behaviors. The strong and fast variations already detected in the Br1 and Br2 regions by the previous spectral analysis seem to be restricted to these two structures while the longer linear variations detected in MC1 are also visible in the MC2 and G0.11-0.11 clouds. A further characterization of the variability trends in adjacent regions shows both increasing and decreasing emissions and highlights an apparent propagation of the signal through the Sgr~A complex. 
\subsection{Schematic view of the variations}
\begin{figure}[h]
	\centering
	\includegraphics[trim = 0mm 7mm 0mm 0mm,clip,width=0.5\textwidth]{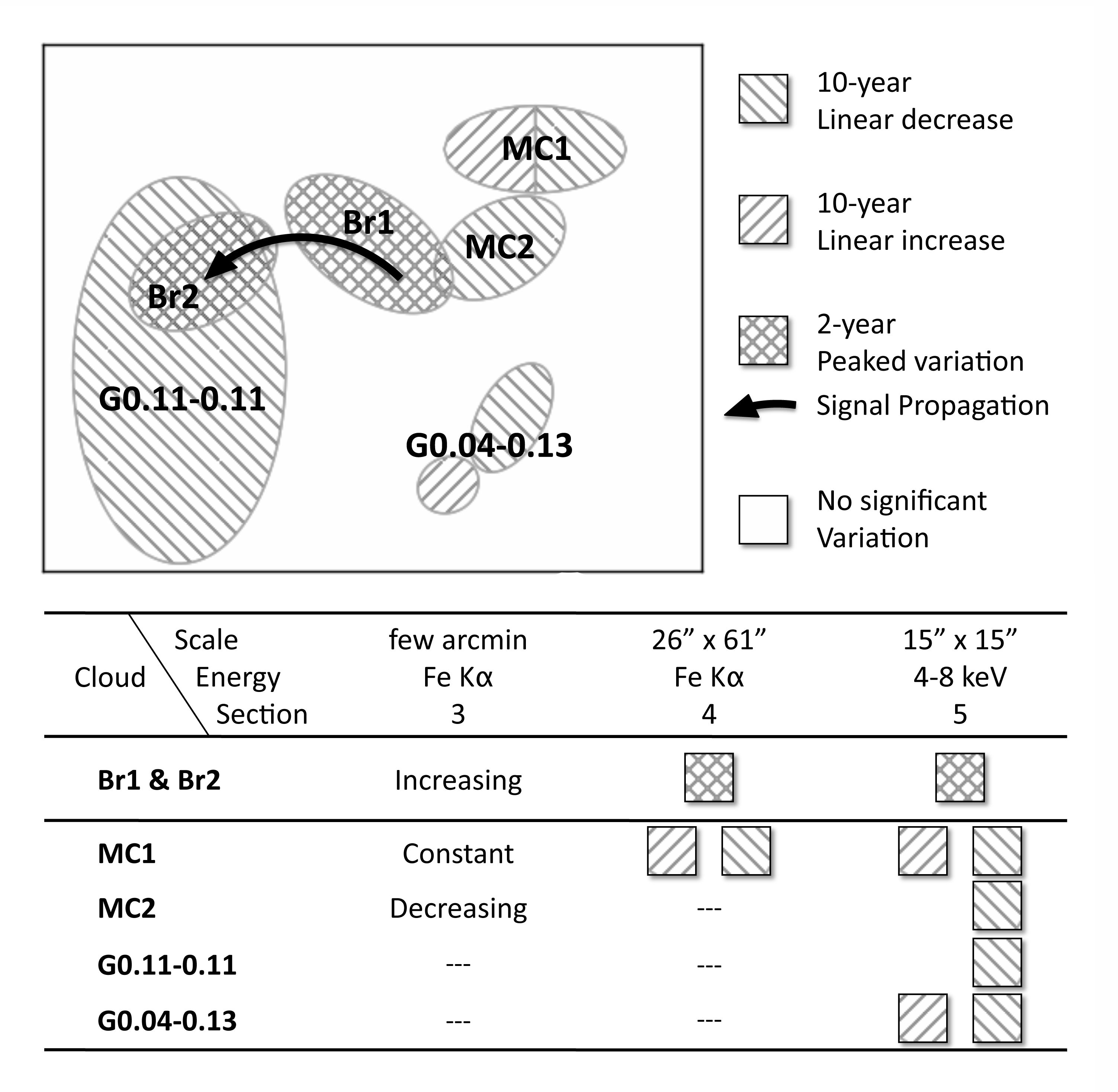}
	\caption{Diagram presenting the main results of the systematic analysis of the 4--8~keV variations. \textit{(Top panel)} spatial representation of the different type of variations. \textit{(Bottom panel)} comparison with the other analyses presented in this work. The different results in section~3 are due to the larger scale analysis there.}
	\label{fig:SchemaSum}
\end{figure}

The results of section 5 are summarized in Figure~\ref{fig:SchemaSum}, showing the varying regions of the Sgr~A complex and their type of variation, and it compares the conclusions of this last analysis to those presented in sections \ref{sec:FeKacloud} and \ref{sec:FeKasmall}. \\\\
From both the \fe\ analysis and the systematic analysis of the 4--8~keV emission at the $15''$~scale, we report a 2-year peaked variation in the Bridge (Br1 and Br2), with a propagation away from \sgr. This behavior is restricted to the Bridge molecular structure. All the other molecular clouds with significant flux changes display linear variations over 10~years, either increasing or decreasing. Therefore there are two different time behaviors in the Sgr~A complex.
%
%
%
%
%
%%%%%%%%%%%%%%%%%%%%%%%%%%%%%%%%%%%% DISCUSSION
\section{Discussion} \label{sec:conclu}
Our 15-arcsecond-scale characterization of the Sgr~A complex reveals an abrupt rise and fall in the Bridge emission with a flux increase of  at least a factor of 10 and lasting no longer than 2~years. This is a strong indication that the illuminating flare for this particular event is either short or has significant temporal substructures.  We also report the detection of much slower linear time variations in both the MC1 and MC2 clouds, linked to morphological changes of the emission in both structures. This clearly indicates a more complex illumination pattern than the constant flux assumed by \citet{ponti2010} to estimate the positions of the clouds along the line of sight in order to assess whether a single long flare could account for all the illumination. Moreover, the slopes of the MC1 and MC2 clouds lightcurves are fully similar to the decreasing trend of G0.11-0.11 variations, also characterized by our arcsecond scale analysis.
This is a hint that at least these three molecular clouds might be witnessing the same event almost simultaneously.\\\\
The two different variation behaviors observed in the Sgr~A complex increase the number of constraints on the global illumination of the region.  We investigate whether all variations can still be explained by one single illuminating event. Since the time behaviour of the cloud emission is the result of a convolution between the illuminating event lightcurve and the cloud structure, the observed difference could indeed be explained by either different cloud morphologies or distinct illuminating episodes having different time signature.\\\\
The following two subsections include calculation results based on equations 1--3 of \citet{sunyaev1998}, which relate the reflected \fe\ flux measured from the cloud to the luminosity of the source. The measured flux depends on the fraction of the total flux intercepted by the cloud (based on size and distance from the source), the cloud column density and iron abundance. The cloud parameters used for this calculation are summarized in table \ref{tab:paramclouds}. 
We point out that the adopted distance between the clouds and \sgr\ combines the projected distance to \sgr\ and the line of sight distance determined by the age of the illuminating event \citep[all fluorescent photons created by an X-ray radiation emitted at a given time are along the same paraboloid,][]{sunyaev1998}. We also assume that the excitation temperature is the same for all clouds and that the clouds have a solar iron abundance and are not self-absorbed at the molecular transition, so the column densities are directly proportional to the antenna temperatures.
\begin{table}[h]
	\centering
	\caption{Average parameters of the five bright clouds.}
	\begin{scriptsize}	
	\begin{tabular*}{0.5\textwidth}{@{\extracolsep{\fill}} l c c c c c}
	\hline \hline
	Parameters & Br1 & Br2 & MC1 & MC2 & G0.11-0.11 \\
	\hline 
 	Min distance to \sgr (pc)\tablefootmark{a} & 17.5 & 23.3 & 11.0 & 13.0 & 25.2\\
 	Flux (10$^{-6}$ ph cm$^{-2}$ s$^{-1}$)\tablefootmark{b} & 5 & 6 & 7 & 6 & 6\\
 	CS antenna temp (K)\tablefootmark{c} & 3 & 2.4 & 1.5 & ... & 1.6 \\ 
 	N$_2$H$^+$ antenna temp (K)\tablefootmark{c} & 1.8 & 1.6 & 0.7 & 0.4 & 0.7 \\
 	Illumination duration (yr)\tablefootmark{d} & $\sim 2$ & $\sim 2$? & $\sim 10$ & $\sim 10$ & $\sim 10$\\
	\hline \hline
	\end{tabular*}
	\label{tab:paramclouds}
	\tablefoot{
	\begin{scriptsize}
	\tablefoottext{a}{Projected distance to \sgr.}
	\tablefoottext{b}{Average maximum 4--8~keV flux measured for each cloud.}
	\tablefoottext{c}{Peak emission value measured in CS \citep{tsuboi1999} and N$_2$H$^+$ \citep{jones2012} maps. The MC2 cloud is not detected in CS maps.}
	\tablefoottext{d}{Illumination period which has been observed. Br2 illumination duration is a prediction since the decrease has not yet been observed.}
		\end{scriptsize}
	}
	\end{scriptsize}	
\end{table}
\subsection{Assumption of one single illuminating event: constraints} \label{sec:oneflare}
Since variations occurring in the Sgr~A complex are spatially organized, it is legitimate to consider that all clouds are illuminated by the same event. In this case the short variation within the Bridge provides the strongest constraint on the profile of the X-ray flare. It implies a strong variation with a duration of less than two years, peaking at least at $10^{39}$~erg~s$^{-1}$ possibly accompanied by a much longer, but ten times weaker, level of activity.
\\\\
To a first approximation, we assume the contribution of any lower level of activity is negligible. Thus, the total duration of the flare is no more than two years and its profile is characterized by a sharp peak. In this case, the longer-term variation seen in the scattered light from all clouds except the Bridge, can only be explained if these clouds have a substantial extent along the line of sight. The extent needed to explain a ten-year illumination with a two-year event depends on the position of the cloud. For instance, the physical line-of-sight extent required for the MC1 or MC2 clouds is between 3~pc for a recent event (50~years ago) and 1.5~pc for an earlier one (300~years) while it is slightly larger for G0.11-0.11 (between 4.5 and 1.5~pc). These values are comparable to the widths of the emission regions measured in the Chandra 6.4~keV flux maps.
\\\\
Nevertheless, if the longer variation seen in the MC1, MC2 and G0.11-0.11 clouds is attributable to a 2-year flare propagating through their interiors, then the illuminated fraction at any given time is less than one third of their total column densities. In particular, the case of a single event implies that all the clouds are along the same paraboloid \citep{sunyaev1998} which means that clouds aligned along the line of sight like Br2 and G0.11-0.11 have to be at the same physical position (despite their different molecular line velocities, Figure~\ref{fig:MOPRA}). Since the observed \fe\ peak intensity of these two clouds is consistent with being the same but with different durations and because they have to be at the same distance from \sgr, the G0.11-0.11 column density should be at least three times larger than the Br2 one. This is more than six times larger than what is implied by the molecular line measurements (table~\ref{tab:paramclouds}). 
The discussion is less straightforward for the two other clouds since MC1 and MC2 could be closer to \sgr\ than the Bridge. This could partly compensate for their lower column density. 
All together, we find that MC1, MC2 and G0.11-0.11 require column densities that are at least three and up to ten times larger than what is inferred from molecular tracers. This excludes a single 2-year flare scenario for the illumination of all these clouds. Adding a longer but ten times lower activity period to this peak does not significantly change this conclusion. 
\\\\
Moreover, as already mentioned, since the linear trends detected in the MC1, MC2 and G0.11-0.11 clouds are very similar, this is a strong hint that this variation pattern is linked to the intrinsic illuminating flare duration rather than to the individual cloud structure. In the case of a single two-year flare, only the cloud density gradient along the line of sight can be invoked to reproduce the linear variation observed on 15-arcsec scale: this means that the three clouds should have a very similar density gradient, which is very unlikely. Once again, adding a ten times lower level of activity before and after the flare peak does not solve this issue.
\\\\
Finally, it is difficult to reconcile all the illumination seen in the CMZ with a single 2-year flare since such an event should be illuminating less than 1\%  of the matter. Moreover, it would constrain the positions of all bright \fe\ clouds to the surface of a thin paraboloid, which also seems very unlikely. Therefore, density, structure and position constraints strongly disfavor the single flare scenario.
\subsection{Towards a scenario with at least two illuminating events}
The Bridge region is illuminated by a 2-year peaked event during which the luminosity of \sgr\ had to be at least 10$^{39}$~erg~s$^{-1}$. The presence of an underlying 10~times fainter component of the illuminated flux having a much longer duration cannot be excluded. 
\\\\
Let suppose the linear variations seen in all the other clouds are due to a single other event.
According to the considerations presented in section \ref{sec:oneflare}, the minimum duration of this second flare can be derived from the cloud linear trend and is found to be substantially longer than the event illuminating the Bridge. Since the reflection time scale can be twice the flare duration when the reflecting clouds are behind the source, we can set a 10-year lower limit on the flare duration (a 5-year increase plus a 5-year decrease). \sgr's past luminosity relative to this longer flare can be estimated using the \nh\ values derived by \citet{ponti2010} and assuming a position along the line of sight. In case of a recent flare (50 years ago), MC1 and MC2 should be very close to the black hole compared to G0.11-0.11. Therefore, the sustained luminosity of \sgr\ during the longer flare should be at least $5\times10^{38}$~erg~s$^{-1}$ and up to few 10$^{39}$~erg~s$^{-1}$ to explain the flux detected respectively in MC1 and G0.11-0.11. This luminosity discrepancy is reduced if we consider older flares. For instance, for a flare fading about 200 years ago, the luminosity of \sgr\ should be of the order of few 10$^{39}$~erg~s$^{-1}$ to explain the fluxes detected in all three clouds. This seems more compatible with the simultaneous and similar decrease observed in these clouds.\\\\
Therefore, two distinct events of similar intensities could account for the variations seen in the Sgr~A complex. Nevertheless, the precise dating of these two flares, their chronological order and the level of emission in between are difficult to assess, and of course we cannot rule out the possibility that there have been more than two flares and that G0.11-0.11 and the MC1 and MC2 clouds have been illuminated by different events, albeit with similar signatures. 
\subsection{Can known Galactic center transients be an alternative to \sgr?}
An important question raised by this two-flare scenario is whether at least one of the events could be due to an X-ray transient rather than to \sgr. The time scale of the event seen in the MC1, MC2 and G0.11-0.11 clouds and in particular the 10-year rise and the 10-year decay observed suggest that this event is not due to one of the typical Galactic center transients \citep{degenaar2012}. By contrast, the sharp and short variations witnessed in the Bridge call \sgr\ into question as a privileged candidate.\\\\
X-ray binaries are legitimate candidates since they can have bursts with luminosities up to $10^{39}$~erg~s$^{-1}$ and some of them are presumably closer to the Bridge molecular cloud than the Galactic center is. Nevertheless, their spectral shape is less favorable for producing the \fe\ line emission than the power-law of photon index  $\Gamma=2$ assumed for \sgr. We compare the luminosity required to produce the \fe\ flux observed in the Bridge ($\rm{L}\,>\,10^{39}$~erg~s$^{-1}$ for one year at the position of \sgr) and the typical luminosities produced by known black hole X-ray binaries \citep{dunn2010}. The distance between the X-ray binary and the cloud is the only free parameter. We find that at the position of the Galactic center, none of these sources \citep[GRS~1915$+$105 and the typical black hole X-ray binaries,][]{dunn2010} is able to explain the emission observed in the Bridge.
At a distance of about 10~pc from the Bridge (twice shorter than the projected distance between the Bridge and \sgr), a source with a one year burst having an intensity comparable to that of GRS~1915$+$105 would account for both the required luminosity and the duration of the observed event. 
Nevertheless, the matter distribution around the position of this hypothetical source would need to be very specific in order to explain the propagation of the echo seen in the Bridge. All other outbursts from typical X-ray binaries (e.g., GX~339--4, GRO~J1655--40, H~1743--322) are excluded since the position of these hypothetical sources should be less than 5~pc away from the Bridge and so they should have been detected in the past 30~years. Therefore, an X-ray binary origin seems very unlikely considering current knowledge of the behavior of Galactic X-ray sources \citep[][and references therein]{dunn2010,degenaar2012}.
\\\\
The recent discovery of the Soft Gamma Repeater (SGR) J1745--29 \citep{kennea2013, mori2013} at only 3~arcsec from \sgr\ \citep{rea2013} suggests that this object could be another plausible candidate for the illuminating event. Indeed SGRs sometimes emit extremely luminous and hard X/$\gamma$-ray bursts. The brightest such burst to date is the 2004 giant flare of SGR~1806--20 that released a total energy of few $10^{46}$~erg in high-energy radiation ($>\,30$~keV) in a fraction of a second \citep{hurley2005}, and a hundred times less energy in the ensuing pulsating tail. In this case only a small fraction of any given cloud would be illuminated at a particular time. Yet, the total energy released in this burst was equivalent to that of a source with a luminosity of 10$^{39}$~erg~s$^{-1}$ active for about one year, making such a giant flare a potentially viable origin for the observed short duration echo. But most of the energy of a giant flare is released in high energy photons beyond 100~keV which do not effectively produce 6.4 keV line emission. To properly evaluate the effective reflected emission produced by such an event occurring at the Galactic center, we used the 2004-flare spectrum of SGR~1806--20 measured by \citet{frederiks2007}: a very hard power-law (photon index $\Gamma = 0.73$) with a cutoff at about 660~keV. We modulated it by the iron cross section for photoelectric absorption and applied the parameters of the Bridge 2011 filament discussed in section~\ref{sec:Bridge}.
Due to the hard spectrum of the source, the computation gives an expected reflected flux in the \fe\ fluorescence line of only about $2\times10^{-8}$~ph~cm$^{-2}$~s$^{-1}$ for the filament while we measure a flux of $3.2\times10^{-6}$~ph~cm$^{-2}$~s$^{-1}$, more than two orders of magnitude higher. For the SGR spectrum measured by \citet{hurley2005} and others the situation is even less favorable. 
A higher density and/or a different geometry for the filament could imply a larger reflected fraction of the illumination and lower the energetic requirements. Yet, the predictions remain one order of magnitude below the measured flux. A flare significantly more energetic than the one produced by SGR~1806--20 in 2004 is thus required to explain the observed emission. Therefore, we conclude that, at the moment, an enhanced past activity of \sgr\ is still the preferred explanation for the outbursts reflected by the molecular clouds of the Sgr~A complex.
%
%
%
%
%
%%%%%%%%%%%%%%%%%%%%%%%%%%%%%%%%%%%%% CONCLU
\section{Conclusion}
Using high spatial resolution data from the Chandra X-ray Observatory, we observed for the first time short and strong variations in the X-ray emission from several clouds in the Sgr A complex. We confirm the general trend on large scales (few arcmin) that indicates that this emission is the reflection of a luminous episode of hard X-ray emission most likely due to the past activity of \sgr. We investigated smaller scale variations and in particular we report:
\vspace{-0.2cm}
\begin{itemize}
	\item The appearance in 2011 of a new bright and thin filament ($0.2\times1$~pc$^2$) in the eastern part of the Bridge, in the direction of the superluminal propagation previously observed \citep{ponti2010}. 
 Based on a reasonable assumption about its density, we conclude that the luminosity of \sgr\ was at least 10$^{39}$~erg~s$^{-1}$ in order to produce such a bright filament.
	\item The first detection of both an increasing and a decreasing phase for the same emitting structures. In particular we identify a 2-year peaked emission that has propagated through the Bridge. This behavior is well characterized by analyses of both the \fe\ line and the 4--8~keV continuum.
	\item The presence of 10-year linear variations in all bright molecular structures of the Sgr~A complex except for the Bridge. In particular we report the first 6.4~keV detection of intrinsic variations in MC1, characterized by a 10-year increase in its eastern side and a decrease in its western side with a similar temporal gradient. The MC2 and G0.11-0.11 clouds show only a 10-year linear decrease.
\end{itemize}
\vspace{-0.2cm}
\noindent
In theory, a 2-year event could account for both the short variation seen in the Bridge and the longer trend detected in all other clouds. Nevertheless, the constraints derived under this hypothesis on the clouds' relative densities are not compatible with values given by molecular tracers and therefore this possibility is excluded. \textit{Thus, the two behaviors characterized for the first time by the present work on the Sgr A complex are likely to be due to the reflection of two distinct past flares of Sgr A$^\star$.} The underlying level of activity of \sgr\ at times surrounding the time of the prominent flares is not well constrained, but is possibly much higher than the current activity level of \sgr. 
Characterizing the frequency of such events along with their durations and their intensities will be fundamental to identifying the physical processes responsible for these past changes in the luminosity of \sgr. They could be due to catastrophic events such as, a transient jet induced by a partial stellar capture \citep{yu2011}, a capture of planets \citep{zubovas2012}, or the accretion of debris produced by tidal interaction of stars \citep{sazonov2012}, but they can also be explained by stochastic variations of the accretion rate due to the emission of clumps by the winds of massive stars orbiting \sgr\ \citep{cuadra2008}.\\\\
The present work emphasizes the importance of characterizing the most rapid variations that manifest themselves on small spatial scales. 
Indeed, we reveal a complex illumination pattern with rapid and fine variations. The timing of these variations is the key for spotlighting the simultaneity in the illuminated cloud behaviors and therefore for constraining the relative positions of the clouds. This aspect is ignored in most other approaches that rather rely on modelling the emission over large regions by averaging their intrinsic behavior.
\\\\
A global scenario relating the detailed variations observed in the Sgr A complex to the other reflection features seen in the CMZ is beyond the scope of this paper. A global cloud modelling and an analysis of the reflection in the overall CMZ will be the subject of a future work.

\begin{acknowledgements} 
The scientific results reported in this article are based on observations made by the Chandra X-ray Observatory and on data obtained from the Chandra Data Archive. This research has made use of software provided by the Chandra X-ray Center (CXC) in the application packages CIAO, ChIPS, and Sherpa. The molecular maps were obtained using the Mopra radio telescope, a part of the Australia Telescope National Facility which is funded by the Commonwealth of Australia for operation as a National Facility managed by CSIRO. The University of New South Wales (UNSW) digital filter bank (the UNSW-MOPS) used for the observations with Mopra was provided with support from the Australian Research Council (ARC), UNSW, Sydney and Monash Universities, as well as the CSIRO. The authors acknowledge the support from the International Space Science Institute to the International Team 216 and the financial support from the UnivEarthS Labex program of Sorbonne Paris Cit\'e (ANR-10-LABX-0023 and ANR-11-IDEX-0005-02). MC acknowledges the Universit\'e Paris Sud 11 for financial support. MM acknowledges support from NASA. GP acknowledges support via an EU Marie Curie Intra-European fellowship under contract no. FP-PEOPLE-2012-IEF-331095. SS acknowledges the Centre National d'Etudes Spatiales (CNES) for financial support. 
\end{acknowledgements}

\bibliographystyle{aa} 
\bibliography{bibSgrAc}

\begin{thebibliography}{45}
\expandafter\ifx\csname natexlab\endcsname\relax\def\natexlab#1{#1}\fi

\bibitem[{{Baganoff} {et~al.}(2003){Baganoff}, {Maeda}, {Morris}, {Bautz},
  {Brandt}, {Cui}, {Doty}, {Feigelson}, {Garmire}, {Pravdo}, {Ricker}, \&
  {Townsley}}]{baganoff2003}
{Baganoff}, F.~K., {Maeda}, Y., {Morris}, M., {et~al.} 2003, \apj, 591, 891

\bibitem[{{Capelli} {et~al.}(2011){Capelli}, {Warwick}, {Cappelluti},
  {Gillessen}, {Predehl}, {Porquet}, \& {Czesla}}]{capelli2011}
{Capelli}, R., {Warwick}, R.~S., {Cappelluti}, N., {et~al.} 2011, \aap, 525, L2

\bibitem[{{Capelli} {et~al.}(2012){Capelli}, {Warwick}, {Porquet}, {Gillessen},
  \& {Predehl}}]{capelli2012}
{Capelli}, R., {Warwick}, R.~S., {Porquet}, D., {Gillessen}, S., \& {Predehl},
  P. 2012, \aap, 545, A35

\bibitem[{{Cuadra} {et~al.}(2008){Cuadra}, {Nayakshin}, \&
  {Martins}}]{cuadra2008}
{Cuadra}, J., {Nayakshin}, S., \& {Martins}, F. 2008, \mnras, 383, 458

\bibitem[{{Degenaar} {et~al.}(2012){Degenaar}, {Wijnands}, {Cackett}, {Homan},
  {in't Zand}, {Kuulkers}, {Maccarone}, \& {van der Klis}}]{degenaar2012}
{Degenaar}, N., {Wijnands}, R., {Cackett}, E.~M., {et~al.} 2012, \aap, 545, A49

\bibitem[{{Dogiel} {et~al.}(2009){Dogiel}, {Cheng}, {Chernyshov}, {Bamba},
  {Ichimura}, {Inoue}, {Ko}, {Kokubun}, {Maeda}, {Mitsuda}, \&
  {Yamasaki}}]{dogiel2009}
{Dogiel}, V., {Cheng}, K.-S., {Chernyshov}, D., {et~al.} 2009, \pasj, 61, 901

\bibitem[{{Dunn} {et~al.}(2010){Dunn}, {Fender}, {K{\"o}rding}, {Belloni}, \&
  {Cabanac}}]{dunn2010}
{Dunn}, R.~J.~H., {Fender}, R.~P., {K{\"o}rding}, E.~G., {Belloni}, T., \&
  {Cabanac}, C. 2010, \mnras, 403, 61

\bibitem[{{Frederiks} {et~al.}(2007){Frederiks}, {Golenetskii}, {Palshin},
  {Aptekar}, {Ilyinskii}, {Oleinik}, {Mazets}, \& {Cline}}]{frederiks2007}
{Frederiks}, D.~D., {Golenetskii}, S.~V., {Palshin}, V.~D., {et~al.} 2007,
  Astronomy Letters, 33, 1

\bibitem[{{Gehrels}(1986)}]{gehrels1986}
{Gehrels}, N. 1986, \apj, 303, 336

\bibitem[{{Ghez} {et~al.}(2008){Ghez}, {Salim}, {Weinberg}, {Lu}, {Do}, {Dunn},
  {Matthews}, {Morris}, {Yelda}, {Becklin}, {Kremenek}, {Milosavljevic}, \&
  {Naiman}}]{ghez2008}
{Ghez}, A.~M., {Salim}, S., {Weinberg}, N.~N., {et~al.} 2008, \apj, 689, 1044

\bibitem[{{Gillessen} {et~al.}(2009){Gillessen}, {Eisenhauer}, {Fritz},
  {Bartko}, {Dodds-Eden}, {Pfuhl}, {Ott}, \& {Genzel}}]{gillessen2009a}
{Gillessen}, S., {Eisenhauer}, F., {Fritz}, T.~K., {et~al.} 2009, \apjl, 707,
  L114

\bibitem[{{Gillessen} {et~al.}(2013){Gillessen}, {Genzel}, {Fritz},
  {Eisenhauer}, {Pfuhl}, {Ott}, {Cuadra}, {Schartmann}, \&
  {Burkert}}]{gillessen2013}
{Gillessen}, S., {Genzel}, R., {Fritz}, T.~K., {et~al.} 2013, \apj, 763, 78

\bibitem[{{Gillessen} {et~al.}(2012){Gillessen}, {Genzel}, {Fritz}, {Quataert},
  {Alig}, {Burkert}, {Cuadra}, {Eisenhauer}, {Pfuhl}, {Dodds-Eden}, {Gammie},
  \& {Ott}}]{gillessen2012}
{Gillessen}, S., {Genzel}, R., {Fritz}, T.~K., {et~al.} 2012, \nat, 481, 51

\bibitem[{{Greene} \& {Ho}(2007)}]{greene2007}
{Greene}, J.~E. \& {Ho}, L.~C. 2007, \apj, 667, 131

\bibitem[{{Hurley} {et~al.}(2005){Hurley}, {Boggs}, {Smith}, {Duncan}, {Lin},
  {Zoglauer}, {Krucker}, {Hurford}, {Hudson}, {Wigger}, {Hajdas}, {Thompson},
  {Mitrofanov}, {Sanin}, {Boynton}, {Fellows}, {von Kienlin}, {Lichti}, {Rau},
  \& {Cline}}]{hurley2005}
{Hurley}, K., {Boggs}, S.~E., {Smith}, D.~M., {et~al.} 2005, \nat, 434, 1098

\bibitem[{{Inui} {et~al.}(2009){Inui}, {Koyama}, {Matsumoto}, \&
  {Tsuru}}]{inui2009}
{Inui}, T., {Koyama}, K., {Matsumoto}, H., \& {Tsuru}, T.~G. 2009, \pasj, 61,
  241

\bibitem[{{Jones} {et~al.}(2012){Jones}, {Burton}, {Cunningham},
  {Requena-Torres}, {Menten}, {Schilke}, {Belloche}, {Leurini},
  {Mart{\'{\i}}n-Pintado}, {Ott}, \& {Walsh}}]{jones2012}
{Jones}, P.~A., {Burton}, M.~G., {Cunningham}, M.~R., {et~al.} 2012, \mnras,
  419, 2961

\bibitem[{{Kennea} {et~al.}(2013){Kennea}, {Burrows}, {Kouveliotou}, {Palmer},
  {G{\"o}{\u g}{\"u}{\c s}}, {Kaneko}, {Evans}, {Degenaar}, {Reynolds},
  {Miller}, {Wijnands}, {Mori}, \& {Gehrels}}]{kennea2013}
{Kennea}, J.~A., {Burrows}, D.~N., {Kouveliotou}, C., {et~al.} 2013, \apjl,
  770, L24

\bibitem[{{Koyama} {et~al.}(2007){Koyama}, {Hyodo}, {Inui}, {Nakajima},
  {Matsumoto}, {Tsuru}, {Takahashi}, {Maeda}, {Yamazaki}, {Murakami},
  {Yamauchi}, {Tsuboi}, {Senda}, {Kataoka}, {Takahashi}, {Holt}, \&
  {Brown}}]{koyama2007}
{Koyama}, K., {Hyodo}, Y., {Inui}, T., {et~al.} 2007, \pasj, 59, 245

\bibitem[{{Koyama} {et~al.}(1996){Koyama}, {Maeda}, {Sonobe}, {Takeshima},
  {Tanaka}, \& {Yamauchi}}]{koyama1996}
{Koyama}, K., {Maeda}, Y., {Sonobe}, T., {et~al.} 1996, \pasj, 48, 249

\bibitem[{{Melia} \& {Falcke}(2001)}]{Melia2001}
{Melia}, F. \& {Falcke}, H. 2001, \araa, 39, 309

\bibitem[{{Mori} {et~al.}(2013){Mori}, {Gotthelf}, {Zhang}, {An}, {Baganoff},
  {Barri{\`e}re}, {Beloborodov}, {Boggs}, {Christensen}, {Craig}, {Dufour},
  {Grefenstette}, {Hailey}, {Harrison}, {Hong}, {Kaspi}, {Kennea}, {Madsen},
  {Markwardt}, {Nynka}, {Stern}, {Tomsick}, \& {Zhang}}]{mori2013}
{Mori}, K., {Gotthelf}, E.~V., {Zhang}, S., {et~al.} 2013, \apjl, 770, L23

\bibitem[{{Morris} \& {Serabyn}(1996)}]{morris1996}
{Morris}, M. \& {Serabyn}, E. 1996, \araa, 34, 645

\bibitem[{{Muno} {et~al.}(2007){Muno}, {Baganoff}, {Brandt}, {Park}, \&
  {Morris}}]{muno2007}
{Muno}, M.~P., {Baganoff}, F.~K., {Brandt}, W.~N., {Park}, S., \& {Morris},
  M.~R. 2007, \apjl, 656, L69

\bibitem[{{Murakami} {et~al.}(2000){Murakami}, {Koyama}, {Sakano}, {Tsujimoto},
  \& {Maeda}}]{murakami2000}
{Murakami}, H., {Koyama}, K., {Sakano}, M., {Tsujimoto}, M., \& {Maeda}, Y.
  2000, \apj, 534, 283

\bibitem[{{Nandra} \& {George}(1994)}]{nandra1994}
{Nandra}, K. \& {George}, I.~M. 1994, \mnras, 267, 974

\bibitem[{{Nobukawa} {et~al.}(2010){Nobukawa}, {Koyama}, {Tsuru}, {Ryu}, \&
  {Tatischeff}}]{nobukawa2010}
{Nobukawa}, M., {Koyama}, K., {Tsuru}, T.~G., {Ryu}, S.~G., \& {Tatischeff}, V.
  2010, \pasj, 62, 423

\bibitem[{{Nobukawa} {et~al.}(2011){Nobukawa}, {Ryu}, {Tsuru}, \&
  {Koyama}}]{nobukawa2011}
{Nobukawa}, M., {Ryu}, S.~G., {Tsuru}, T.~G., \& {Koyama}, K. 2011, \apjl, 739,
  L52

\bibitem[{{Nowak} {et~al.}(2012){Nowak}, {Neilsen}, {Markoff}, {Baganoff},
  {Porquet}, {Grosso}, {Levin}, {Houck}, {Eckart}, {Falcke}, {Ji}, {Miller}, \&
  {Wang}}]{nowak2012}
{Nowak}, M.~A., {Neilsen}, J., {Markoff}, S.~B., {et~al.} 2012, \apj, 759, 95

\bibitem[{{Park} {et~al.}(2004){Park}, {Muno}, {Baganoff}, {Maeda}, {Morris},
  {Howard}, {Bautz}, \& {Garmire}}]{park2004}
{Park}, S., {Muno}, M.~P., {Baganoff}, F.~K., {et~al.} 2004, \apj, 603, 548

\bibitem[{{Ponti} {et~al.}(2013){Ponti}, {Morris}, {Terrier}, \&
  {Goldwurm}}]{ponti2012}
{Ponti}, G., {Morris}, M.~R., {Terrier}, R., \& {Goldwurm}, A. 2013, in
  Advances in Solid State Physics, Vol.~34, Cosmic Rays in Star-Forming
  Environments, ed. D.~F. {Torres} \& O.~{Reimer}, 331

\bibitem[{{Ponti} {et~al.}(2010){Ponti}, {Terrier}, {Goldwurm}, {Belanger}, \&
  {Trap}}]{ponti2010}
{Ponti}, G., {Terrier}, R., {Goldwurm}, A., {Belanger}, G., \& {Trap}, G. 2010,
  \apj, 714, 732

\bibitem[{{Rea} {et~al.}(2013){Rea}, {Esposito}, {Israel}, {Papitto}, {Tiengo},
  {Baganoff}, {Haggard}, {Mereghetti}, {Burgay}, {Possenti}, \&
  {Zane}}]{rea2013}
{Rea}, N., {Esposito}, P., {Israel}, G.~L., {et~al.} 2013, ATel, 5032

\bibitem[{{Revnivtsev} {et~al.}(2004){Revnivtsev}, {Churazov}, {Sazonov},
  {Sunyaev}, {Lutovinov}, {Gilfanov}, {Vikhlinin}, {Shtykovsky}, \&
  {Pavlinsky}}]{revnivtsev2004}
{Revnivtsev}, M.~G., {Churazov}, E.~M., {Sazonov}, S.~Y., {et~al.} 2004, \aap,
  425, L49

\bibitem[{{Ryu} {et~al.}(2012){Ryu}, {Nobukawa}, {Nakashima}, {Tsuru},
  {Koyama}, \& {Uchiyama}}]{ryu2012}
{Ryu}, S.~G., {Nobukawa}, M., {Nakashima}, S., {et~al.} 2012, ArXiv: 1211.4529

\bibitem[{{Sazonov} {et~al.}(2012){Sazonov}, {Sunyaev}, \&
  {Revnivtsev}}]{sazonov2012}
{Sazonov}, S., {Sunyaev}, R., \& {Revnivtsev}, M. 2012, \mnras, 420, 388

\bibitem[{{Su} {et~al.}(2010){Su}, {Slatyer}, \& {Finkbeiner}}]{su2010}
{Su}, M., {Slatyer}, T.~R., \& {Finkbeiner}, D.~P. 2010, \apj, 724, 1044

\bibitem[{{Sunyaev} \& {Churazov}(1998)}]{sunyaev1998}
{Sunyaev}, R. \& {Churazov}, E. 1998, \mnras, 297, 1279

\bibitem[{{Sunyaev} {et~al.}(1993){Sunyaev}, {Markevitch}, \&
  {Pavlinsky}}]{sunyaev1993}
{Sunyaev}, R.~A., {Markevitch}, M., \& {Pavlinsky}, M. 1993, \apj, 407, 606

\bibitem[{{Tatischeff} {et~al.}(2012){Tatischeff}, {Decourchelle}, \&
  {Maurin}}]{tatischeff2012}
{Tatischeff}, V., {Decourchelle}, A., \& {Maurin}, G. 2012, \aap, 546, A88

\bibitem[{{Terrier} {et~al.}(2010){Terrier}, {Ponti}, {B{\'e}langer},
  {Decourchelle}, {Tatischeff}, {Goldwurm}, {Trap}, {Morris}, \&
  {Warwick}}]{terrier2010}
{Terrier}, R., {Ponti}, G., {B{\'e}langer}, G., {et~al.} 2010, \apj, 719, 143

\bibitem[{{Tsuboi} {et~al.}(1999){Tsuboi}, {Handa}, \& {Ukita}}]{tsuboi1999}
{Tsuboi}, M., {Handa}, T., \& {Ukita}, N. 1999, \apjs, 120, 1

\bibitem[{{Yu} {et~al.}(2011){Yu}, {Cheng}, {Chernyshov}, \& {Dogiel}}]{yu2011}
{Yu}, Y.-W., {Cheng}, K.~S., {Chernyshov}, D.~O., \& {Dogiel}, V.~A. 2011,
  \mnras, 411, 2002

\bibitem[{{Yusef-Zadeh} {et~al.}(2002){Yusef-Zadeh}, {Law}, \&
  {Wardle}}]{yusef2002}
{Yusef-Zadeh}, F., {Law}, C., \& {Wardle}, M. 2002, \apjl, 568, L121

\bibitem[{{Zubovas} {et~al.}(2012){Zubovas}, {Nayakshin}, \&
  {Markoff}}]{zubovas2012}
{Zubovas}, K., {Nayakshin}, S., \& {Markoff}, S. 2012, \mnras, 421, 1315

\end{thebibliography}

\end{document}